\documentclass[a4paper,12pt]{article}
\pdfoutput=1 
\usepackage[margin=1in,includefoot]{geometry}
\usepackage{amsmath,amssymb,amsthm,eufrak} 
\usepackage[pdfstartview=FitH,pdfpagemode=None]{hyperref}
\usepackage[noadjust]{cite} 
\usepackage{authblk} 
\renewcommand\Affilfont{\itshape\small}
\usepackage{xfrac} 
\usepackage{cancel} 
\usepackage{comment} 
\usepackage{empheq}

\usepackage[sans]{dsfont} 
\usepackage{multirow}
\usepackage{array}     
\usepackage{makecell}  

\DeclareFontFamily{U}{dsss}{}
\DeclareFontShape{U}{dsss}{m}{n}{
  <-10>   s*[1] dsss8
  <10-12> s*[1] dsss10
  <12->   s*[1] dsss12
}{}
\DeclareMathAlphabet{\mathds}{U}{dsss}{m}{n}

\usepackage[scaled=1.2]{dsserif} 

\usepackage{color} 
\usepackage{soul}
\numberwithin{equation}{section} 

\def\be{\begin{equation}}
\def\ee{\end{equation}}
\def\bea{\begin{eqnarray}}
\def\eea{\end{eqnarray}}

\newcommand{\Tr}{{\rm Tr}}

\newcommand{\wg }{\wedge}
\renewcommand{\to}{\rightarrow}

\def\nb{\nonumber}
\def\CP{\mathds C \text P}

\def\pt{\hspace{1pt}}

\newcommand{\rmd}{\,\mathrm{d}}
\definecolor{cardinal}{rgb}{0.6,0,0}
\definecolor{darkgreen}{rgb}{0,0.5,0}
\definecolor{golden}{rgb}{0.92, 0.7, 0}
\definecolor{midnight}{rgb}{0, 0, 0.5}
\definecolor{darkblue}{rgb}{0.2, 0, 0.8}

\newcommand{\ie}{i.e.,\ }
\newcommand{\eg}{e.g.,\ }

\def\({\left (}
\def\){\right )}

\let\benn\[
\let\eenn\]

\def\[{\left [}
\def\]{\right ]}

\let\oldlgraf\{ 
\renewcommand{\{}{\left \oldlgraf}

\let\oldrgraf\}
\renewcommand{\}}{\right \oldrgraf}

\title{\bfseries\LARGE Hagedorn temperature from holography:\protect\\rotating and charged systems}

\author[a]{Francesco Bigazzi}
\author[b,c]{Tommaso Canneti}
\author[d]{\authorcr Federico Castellani}
\author[a,e]{Aldo L.~Cotrone}

\renewcommand\Affilfont{\itshape\small}

\affil[a]{INFN, Sezione di Firenze%
\protect\\ Via G.~Sansone 1, I-50019 Sesto Fiorentino (Firenze), Italy.}
\affil[b]{Dipartimento di Fisica, Universit\`a di Torino %
\protect\\ Via Pietro Giuria 1; 10125 Torino, Italy.}
\affil[c]{Istituto Nazionale di Fisica Nucleare, Sezione di Torino %
\protect\\ Via Pietro Giuria 1; 10125 Torino, Italy.}
\affil[d]{Institut fur Physik, Humboldt-Universitat zu Berlin and IRIS Adlershof,%
\protect\\ Zum Gro\ss en Windkanal 2, 12489 Berlin, Germany.}
\affil[e]{Dipartimento di Fisica e Astronomia, Universit\`a di Firenze%
\protect\\  Via G. Sansone 1, I-50019 Sesto Fiorentino (Firenze), Italy.}

\date{\small bigazzi@fi.infn.it, tommaso.canneti@unito.it,%
\authorcr federico.castellani@physik.hu-berlin.de, cotrone@fi.infn.it}
\begin{document}

%
\maketitle

\begin{abstract}
Using the string/field theory correspondence as a tool, we study the dependence of the Hagedorn temperature of strongly coupled, planar gauge theories on angular velocities and chemical potentials for $U(1)$ global currents. Our results are obtained from an interplay of world-sheet semiclassical quantization methods and target space low energy effective ones. The Hagedorn temperature is given as an expansion in the inverse (large) 't Hooft coupling limit. Working to quadratic order in the fluctuations, the world-sheet analysis provides the leading, the next-to-leading and part of the next-to-next-to leading order terms. In particular, it captures the NNLO $\log 2$ terms which are due to the contributions of non-zero modes to the zero-point energy. These modes are not accounted for by the effective approach, within which they have to be included by hand using suitable educated guesses. On the other hand, the effective methods allow to compute in a relatively easy way the missing NNLO pieces (and even further subleading corrections), which could be accounted for from the world-sheet approach only working at quartic (or higher) order in the fluctuations. We present general results and several examples in various dimensions, including both confining models and conformal field theories on spheres (such as ${\cal N}=4$ SYM and ABJM) dual to string theories on global $AdS$ spaces. 

\end{abstract}


\newpage
\tableofcontents


\section{Introduction}
The partition function of quantum theories whose density of states grows exponentially with the energy, tends to diverge when the temperature approaches the so-called Hagedorn temperature $T_H$ from below. Relevant examples include some confining gauge theories such as Yang-Mills, conformal field theories (CFTs) on compact manifolds and string theories. 

Computing $T_H$ from first principles in strongly coupled quantum field theories (QFTs) is extremely challenging in general. In recent years relevant results have been obtained, in the planar regime, for ${\cal N}=4$ supersymmetric Yang-Mills (SYM) on $S^3$ \cite{Harmark:2021qma,Ekhammar:2023glu} and for the ABJM theory on $S^2$ \cite{Ekhammar:2023cuj}, using quantum spectral curve (QSC) methods. In both cases, $T_H$ has been extracted numerically up to several subleading orders in an expansion in powers of the inverse 't Hooft coupling $1/\lambda$. Indicative results for Yang-Mills theories have also been obtained using lattice methods (see, \eg \cite{Caselle:2015tza}), although the latter still miss first principle derivations of $T_H$. 

In order to partially fill the gap, the computation of the Hagedorn temperature of strongly coupled planar quantum field theories, admitting a dual holographic description in terms of string theory models on curved backgrounds, has been the subject of several works in the literature. In the past few years relevant progresses have been made and $T_H$ has been computed, using holographic methods, up to several subleading orders in the $\alpha'$ (or $1/\lambda$) expansion, for a large class of theories \cite{Urbach:2022xzw,Bigazzi:2022gal,Urbach:2023npi,Ekhammar:2023glu,Bigazzi:2023oqm,Ekhammar:2023cuj,Bigazzi:2023hxt, Harmark:2024ioq,Bigazzi:2024biz,Bigazzi:2024sjy,Canneti:2025rsp}. The class includes both strictly confining models and CFTs on spheres. Remarkably, in the above mentioned ${\cal N}=4$ SYM and ABJM cases, the string theory results perfectly agree with the quantum spectral curve ones up to next-to-next-to-next-to leading order (NNNLO) in the strong coupling expansion \cite{Ekhammar:2023glu,Bigazzi:2023hxt,Ekhammar:2023cuj}. The origin of these successes and of the reached precision lies in the interplay of semiclassical world-sheet methods and effective approaches. 

The world-sheet approach, developed in \cite{Bigazzi:2022gal,Bigazzi:2023oqm, Bigazzi:2023hxt,Bigazzi:2024biz,Bigazzi:2024sjy,Canneti:2025rsp}, consists on computing the spectrum of the quadratic fluctuations of a string around a classical configuration with unit winding number around the Euclidean time direction. The string is chosen to probe an asymptotic region of the gravity solution holographically dual to the quantum field theory under scrutiny. The region corresponds to the QFT infrared regime. The Hagedorn temperature is then obtained as the one above which the ground state of the winding string becomes tachyonic \cite{Atick:1988si}. 

The ground state corresponds, in target space, to a complex scalar field usually referred to as the “thermal scalar". When the temperature $T$ approaches $T_H$ from below, the mass of the field tends to zero and thus the thermal scalar has to be included in the string theory low energy effective action. The contribution of the thermal scalar to the latter can be deduced and serves as a complementary starting point to access $T_H$, along the lines of the Horowitz-Polchinski construction \cite{Horowitz:1997jc}. 

The interplay of the two approaches reveals to be crucial to capture the full NNLO and NNNLO terms in the strong coupling expansion of $T_H$. In particular, the semiclassical world-sheet approach, to quadratic order in the fluctuations, allows to compute the LO, the NLO and part of the NNLO term. In particular, it captures the terms proportional to $\log 2$ in the latter. These arise from the contribution of non-zero modes to the world-sheet zero-point energy. As such, these are in turn missing from the target space effective approach, which instead captures, in a relatively simple way, the missing NNLO terms (on top of the LO and NLO ones). The latter could also be computed from the world-sheet, at the price of pushing the semiclassical approach to quartic order, something which can be quite demanding in general. 

A general proposal on how to include the missing $\log 2$ terms in the effective approach has been suggested in \cite{Harmark:2024ioq} and it has been recently tested in \cite{Ekhammar:2025efc} to compute (to NNNLO) the Hagedorn temperature of ${\cal N}=4$ SYM on $S^3$ both at finite chemical potential for a $U(1)$ subgroup of the R-symmetry group, and at finite angular velocities in $S^3$. The work in \cite{Ekhammar:2025efc} follows up a previous paper \cite{Seitz:2025wpc} where rotating strings in flat space or in global $AdS$ have been considered, and the related angular-velocity-dependent Hagedorn temperature has been computed (up to NLO in the $AdS$ cases). The string theory results collected in \cite{Ekhammar:2025efc}, complementing the analysis of \cite{Seitz:2025wpc}, turned out to match the QSC ones at least to NNLO (at subleading order, in fact, the QSC results turned out to be numerically unstable).  

In view of the above mentioned successes, it is important to see if and how the first principle world-sheet predictions on the Hagedorn temperature (in particular those on the above mentioned $\log 2$ terms) are modified by chemical potentials and angular velocities. It is also important to extend the analysis of \cite{Seitz:2025wpc} and \cite{Ekhammar:2025efc} to more general classes of quantum field theories with known holographic dual descriptions. These are precisely the main aims of the present work. 

We thus compute, to NNLO (and in the global $AdS$ cases to NNNLO) in the strong coupling expansion, the Hagedorn temperature of a large class of QFTs, at finite values of the ``chemical potentials" that couple either to global $U(1)$ charges or to angular momenta.\footnote{To include both types of charges, we use the quotation marks in writing \emph{``chemical potentials''}.} Our analysis includes both (quiver) CFTs compactified on $(d-1)$-dimensional spheres and strictly confining gauge theories. In the first case the dual holographic backgrounds feature a $(d + 1)$-dimensional Anti de Sitter sector in global coordinates (global-$AdS_{d+1}$), times a transverse $9-d$ dimensional space $\mathcal M$.\footnote{These are sometimes exact solutions of the Einstein field equations to all orders in the string length.} 

Our analysis provides a first-principle confirmation, to NNNLO, of the holographic results for $\mathcal N = 4$ SYM on $\mathds R \times S^3$, obtained in \cite{Ekhammar:2025efc} using the conjectures of \cite{Harmark:2024ioq}. Moreover we present novel predictions, to NNNLO, for the Hagedorn temperature of spinning (or charged) ABJM on $\mathds R \times S^2$, for theories with global $AdS_3$ duals, and for quiver CFTs dual to global $AdS_5\times X_5$ backgrounds, with $X_5$ being a Sasaki-Einstein space, with a twist along the Reeb vector direction.  Further results are provided for more general confining holographic models, whose master example is Witten's holographic Yang-Mills (WYM) \cite{Witten:1998zw}. 

This work is organized as follows. In section \ref{sec:gen_proposal}, using semiclassical world-sheet methods, we present a general equation for the Hagedorn temperature which includes the aforementioned NNLO $\log 2$ terms and takes into account the twists in the world-sheet frequencies possibly induced by the rotations. In section \ref{Sec:confining_bg} we solve the equation for a large class of examples, including holographic confining gauge theories and CFTs on spheres. In section \ref{sec:effective} we complement the previous world-sheet results with the thermal scalar approach, which allows us to obtain the complete expression for the Hagedorn temperature of spinning or charged models to NNNLO in the strong coupling expansion. We present some concluding remarks in section \ref{sec:conclu} and add further details in the appendices.  

\section{A general equation from the world-sheet}
    \label{sec:gen_proposal}

In this section we will extend the world-sheet results on the Hagedorn temperature of holographic confining gauge theories (and of CFTs on compact spaces) obtained in \cite{Bigazzi:2023hxt,Bigazzi:2024biz, Bigazzi:2024bfq,Canneti:2025cos}, to setups with finite chemical potentials or finite angular velocities. 

The starting point in the above cited papers was the quadratic world-sheet sigma-model for non-rotating Type II Green-Schwarz (GS) superstrings on curved backgrounds dual to the aforementioned classes of gauge theories. The sigma-model was obtained by considering the fluctuations around a classical configuration for a string sitting at the origin of the holographic radial coordinate (corresponding to the deep infra-red regime of the dual field theory) and winding once along the compact Euclidean time direction. The extrapolation to the Hagedorn regime was realized by focusing on fluctuations around a limiting classical configuration with zero momentum\footnote{In this section we write the classical string configuration with capital letters as customary, while in the following sections we are going to switch to lowercase letters for simplicity.}
\be\label{Hagreg}
X^0=\pm \frac{\beta_H}{2\pi} \sigma \, , \quad \vec X = 0 \, ,
\ee
where $\beta_H=1/T_H$ denotes the inverse Hagedorn temperature, $\sigma\in [0,2\pi]$ is the compact world-sheet coordinate and “0” stands for the (compact) time direction of the target space. Furthermore, $\vec X$ symbolically collects all the other directions. 

As in the Atick-Witten \cite{Atick:1988si} flat-space approach,  the Hagedorn temperature was obtained from the quantum mass-shell condition of the quadratic sigma-model, as the temperature at which the ground state of the winding string becomes massless (and above which it becomes tachyonic). This translates in the following implicit equation for $T_H$
\be\label{implicitbetaH}
\frac{g_{00}(0)}{\alpha'} \frac{\beta^2_H}{8\pi^2} = \Delta(\beta_H) + \Delta \mathcal E(\beta_H) \, .
\ee
Here, $g_{00}$ is the ``$00$'' component of the background metric computed on the classical solution \eqref{Hagreg}. 
For static, isotropic, holographic backgrounds, $g_{00}$ depends only on the holographic radial coordinate: hence in the formula above it has to be computed at the value of the latter corresponding to the IR regime of the dual field theory. In general, for holographic confining theories, it can be expressed as
\be
\label{g000}
g_{00}(0) = 2\pi\alpha' T_s \, ,
\ee
where $T_s$ is the confining string tension\footnote{For notational convenience we use the improper terminology ``\emph{string tension}'' for CFTs as well.} and $\alpha'$ is the string length squared. 

A crucial ingredient of equation \eqref{implicitbetaH} is the zero-point energy $\Delta$ of the quadratic world-sheet sigma model at the Hagedorn point. We will come back to this term in a moment. The other term,
\be\label{gendeltaE}
\Delta \mathcal E(\beta_H) = k_2 \, \frac{\beta_H^2}{4\pi^2} + k_3 \, \frac{\beta_H^3}{8\pi^3} + \mathcal{O}(\beta_H^4)\,,
\ee
collects all the contributions which arise going beyond quadratic order in the string fluctuations. As we have partially anticipated in the Introduction, and as we will also see in section \ref{sec:effective} (see in particular formula (\ref{effk2k3})), the coefficients $k_2,k_3$ in $\Delta \mathcal E$ are thus usually computed within the easier-to-handle thermal scalar effective approach, although they should be in principle accessible from the world-sheet too.
\footnote{For the time being, the computation of $k_2$ from the quartic world-sheet theory is available only for strings at rest in a global-$AdS$ background \cite{Bigazzi:2024biz}. We leave the analysis of the general case for future work.}

Remarkably, the zero-point energy $\Delta$ can be expressed as the following power series
\be\label{oldDelta}
\Delta = 1 - \frac12 \sum_{b=1}^8 \mu_b + \log 2 \sum_{b=1}^8 \mu^2_b + \ldots \, ,
\ee
where $\mu_b$ is the $\beta_H$-dependent mass of the $b$-th physical world-sheet boson. The first two terms in \eqref{oldDelta} arise as contributions of the world-sheet zero-modes: as such, they can also be recovered from the target space thermal scalar effective approach. The subleading $\log 2$ term, instead, is due to world-sheet non-zero modes, which are not captured by the thermal scalar effective action. Hence, while from the world-sheet they are computed from first principles, in the effective approach they have to be added by hand under suitable general assumptions \cite{Harmark:2024ioq}. Further subleading corrections in the expansion of $\Delta$ arise at $\mathcal{O}(\mu_b^4)$, so that the $\mathcal{O}(\beta_H^3)$ term in equation \eqref{implicitbetaH} is fully encoded by $k_3$ in \eqref{gendeltaE}.

The parameters $\mu_b$ can be computed as the eigenvalues of the bosonic mass matrix $\mathcal M_b$, whose components are\footnote{For the sake of simplicity, we set the Kalb-Ramond field of the supergravity background to zero.}
\be\label{bosmassmatr}
\( \mathcal M_b \)^m{}_n = -\eta^{\alpha\beta} \partial_\alpha X^p \partial_\beta X^q R^m{}_{p n q}(X) \, , \quad \forall m, \, n, \,p, \, q \, , \quad \eta = \text{diag}\{-,+\} \, .
\ee
Here, Greek letters $\alpha,\beta$, refer to the world-sheet coordinates $(\tau,\sigma)$, while $R^m{}_{p n q}$ are the components of the Riemann tensor of the target space (to be computed on the classical string configuration). Extrapolating to the limiting Hagedorn configuration in \eqref{Hagreg}, it is thus clear how the world-sheet masses depend on $\beta_H$. Moreover, notice that
\be\label{bosonictrace}
\Tr \( \mathcal M_b \) = - \eta^{\alpha\beta} \partial_\alpha X^p \partial_\beta X^q R_{pq}(X) = \sum_{b=1}^8 \mu^2_b \, , \quad R_{pq}(X) = R^m{}_{p m q}(X) \, ,
\ee
provides the sum of the squared masses of the eight transverse bosonic modes (the longitudinal directions being massless). 

Crucially, the knowledge of the masses $\mu_f$ of the fermionic world-sheet modes is not required to compute the $\log 2$ term in equation \eqref{oldDelta}. This is a consequence of the conformal invariance on the world-sheet which implies the mass-matching condition $\sum_b \mu_b^2=\sum_f \mu_f^2$.

Solving \eqref{implicitbetaH} and thus computing $T_H$ in an expansion in $\alpha'$, using an interplay of world-sheet and target space effective methods, it has been possible to provide spectacular precision tests of holography \cite{Bigazzi:2023hxt,Ekhammar:2023glu, Ekhammar:2023cuj}, whenever the QFT results at strong coupling were available via Quantum Spectral Curve methods~\cite{Harmark:2021qma, Ekhammar:2023glu, Ekhammar:2023cuj}. 

Let us now see how the above results can be extended when the QFTs under scrutiny display finite ``chemical potentials" that couple either to $U(1)$ global charges (for instance R-charges in supersymmetric theories) or to the spins of the operators. In the dual ten-dimensional gravity solutions, turning on such chemical potentials would amount to twisting certain angular directions (in a suitable cylindrical coordinate system) with the time direction: starting from static backgrounds, this generates stationary rotating solutions. From the point of view of the classical string configuration adopted to compute $T_H$, these twists would correspond to rotations in suitably chosen planes \cite{Seitz:2025wpc}. The same setup might also be designed with a static string probing the twisted background (see also appendix \ref{app:flatspace}).


Now, proper motion of the string could give rise to the obstructions discussed in \cite{Bigazzi:2024biz}, which might prevent us from simplifying the equations of motion for the world-sheet fluctuations. To fix ideas, let us consider linear embedding maps and vanishing Kalb-Ramond field. Then, in the conformal gauge, the on-shell bosonic fluctuations collected in the vector field $\zeta$ are solutions of
\be
-\eta^{\alpha\beta} D_\alpha D_\beta \, \zeta + \mathcal M_b \cdot \zeta  = 0 \, ,
\ee
$D_\alpha$ being the covariant derivative defined using the Levi-Civita connection. The above equation translates into 
\be
-\eta^{\alpha\beta} \partial_\alpha \partial_\beta \, \zeta + 2 \hspace{1pt} i \,  \eta^{\alpha\beta} A_\alpha \hspace{-2pt} \cdot \partial_\beta \, \zeta + \(\mathcal M_b + i \, \eta^{\alpha\beta} \partial_\alpha A_\beta + \eta^{\alpha\beta} A_\alpha \hspace{-2pt}\cdot \hspace{-2pt} A_\beta \) \cdot \zeta = 0 \, ,
\ee
where we introduced $A_\alpha$ as the matrix having components
\be
\label{AChristoffels}
\(A_\alpha\)^k{}_l = i \hspace{1pt} \partial_\alpha X^p \, \Gamma^k{}_{pl}(X) \,,
\ee
fixed by the pullback on the world-sheet of the Christoffel symbols of the target space (evaluated on the classical string configuration).\footnote{See also appendix \ref{app_geom}.}

In the following, we will focus on stationary supergravity backgrounds in the Hagedorn regime. The limiting configuration in \eqref{Hagreg} can be understood as being written in the comoving frame of the string. There, the bosonic equations of motion reduce to
\be\label{Hagboseom}
(\partial^2_\tau -\partial^2_\sigma) \, \zeta + 2 \hspace{1pt} i \,  A  \hspace{-2pt} \cdot \partial_\sigma \, \zeta + \(\mathcal M_b + A^2 \) \cdot \zeta = 0 \, , \quad A^k{}_l \equiv i \hspace{1pt} \frac{\beta_H}{2\pi} \Gamma^k{}_{0l} \, .
\ee
Let $U$ be the square matrix such that
\be
\mathcal A = U^{-1} A U \, , \quad \Omega = U^{-1} \(\mathcal M_b + A^2 \) U \, ,
\ee
are both diagonal matrices.\footnote{In principle, it is not immediately obvious that the connection and the mass matrix commute with each other at the Hagedorn point. Nevertheless, such a property holds in all the cases of interest analyzed in the following sections. It is thus possible to simultaneously diagonalize them.
}
Then, the above equations simplify to
\be\label{diagonalizedgeneom}
(\partial^2_\tau -\partial^2_\sigma) \, \xi + 2 \hspace{1pt} i \,  \mathcal A \hspace{-2pt} \cdot \partial_\sigma \, \xi + \Omega \cdot \xi = 0 \, , \quad \xi = U^{-1} \cdot\zeta  \, .
\ee
Here, $U$ is assumed to be constant.

The solutions of the diagonalized system can be parameterized as
\begin{subequations}\label{genbosonicmodes}
\be
\xi^b = \xi^b_0 + \Xi^b \, ,
\ee
where
\begin{align}
&\xi^b_0 = i \sqrt{\frac{\alpha'}{2 \, \omega_{0b}}} \, \Bigl( a^b e^{- i \hspace{1pt} \omega_{0b} \hspace{1pt} \tau} - (a^b)^\dagger e^{+i \hspace{1pt} \omega_{0b} \tau}\Bigr) \, , \\[1ex]
& \Xi^b = i \sqrt{\frac{\alpha'}{2}} \sum_{n \neq 0} \frac{1}{\omega_{nb}}  \Bigl(\alpha^b_n e^{- i \hspace{1pt} n \hspace{1pt} \sigma} + \tilde\alpha^b_n e^{+ i \hspace{1pt} n \hspace{1pt} \sigma}\Bigr) \, e^{-i \hspace{1pt} \omega_{nb} \tau} \, ,
\end{align}
and
\be\label{genericshiftedharmonics}
\omega_{0b} = \sqrt{\mu^2_b + \nu^2_b} \, , \quad \omega_{nb} = \text{sgn}(n) \sqrt{(n+\nu_b)^2+\mu^2_b} \, , \quad \forall \,n \in \mathds Z \setminus \{0\} \, .
\ee
\end{subequations}
Here, we denoted as $\omega^2_{0b}$ and $\nu_b$ the eigenvalues of $\Omega$ and $A$ respectively. If non-trivial, the latter shift the standard harmonics as indicated in the above relations. In appendix \ref{app:flatspace}, we show how this structure arises naturally in the case of rotating strings, with the $\nu_b$ being related to (possibly imaginary\footnote{\label{imaginaryfoot}Twisting an angular direction $\phi$ with Minkowski time realizes into the map $\phi\mapsto\phi-\Omega\pt t$. Here, $\Omega$ is the physical angular velocity of the string. In this work, we deal with Euclidean time. The Wick-rotation $t\mapsto-i\pt t$ affects the above twist as $\phi\mapsto\phi+i\pt\Omega\pt t$. We can choose to work with either real angular velocities $\Omega$ or imaginary angular velocities $\Omega=i\pt\omega$, $\omega$ being real. Whenever we discuss real or imaginary angular velocities, we will refer to $\Omega$ in this notation. See \cite{Seitz:2025wpc} for a detailed discussion on the ensembles where a Hagedorn behavior for rotating strings is expected.}) angular velocities. Similarly, for the frequencies of the fermionic world-sheet modes, we would have
\be
\omega_{fr} = \text{sgn}(r) \sqrt{(r+\nu_f)^2+\mu_f^2} \, , \quad \forall \, r \in \mathds Z_{+1/2} \, ,
\ee
where $\mathds Z_{+1/2}$ stands for the set of all possible half-integers and $f$ runs over all the possible fermionic physical world-sheet modes having mass $\mu_f$. In the following, we will assume that
\be\label{shiftmatch}
\Tr A^2 = \sum_b \nu^2_b = \sum_f \nu^2_f \, .
\ee
We will refer to this relation as \emph{shift-matching condition}. It is not crucial for the consistency of the world-sheet sigma-model itself. Anyway, it can be shown that it holds under the current assumptions \cite{toappear}.

Notice that, under any diffeomorphism keeping the temporal direction unaffected, $A$ transforms as a mixed two-rank tensor, namely
\be
A' = J \cdot A \cdot J^{-1} \, ,
\ee
$J$ being the Jacobian matrix of the transformation. This automatically implies the invariance of the characteristic polynomial of $A$, i.e.
\be
\mathcal P(\lambda) = \det \(A - \lambda \, \mathbb 1\) \, ,
\ee
under any such change of coordinates. On the other hand, $\mathcal M_b$ features a manifest tensorial structure. All in all, we conclude that the harmonic expansion in \eqref{genbosonicmodes} holds in any frame whose temporal direction is classically parameterized as in \eqref{Hagreg}.

Thus, we are now allowed to define the generic zero-point energy of a closed type II superstring winding once the compact temporal direction at the Hagedorn point, and probing a twisted ten-dimensional background dual to a confining gauge theory at finite temperature and finite ``chemical potentials" as
\be\label{genDelta}
\Delta = - \frac12 \sum_{b=1}^8 \sum_{n\in\mathds Z} \sqrt{(n+\nu_b)^2+\mu^2_b} + \frac12 \sum_{f=1}^8 \sum_{r\in\mathds Z_{+1/2}} \sqrt{(r+\nu_f)^2+\mu^2_f} \, .
\ee
The sums run over all the possible bosonic ($b$) and fermionic ($f$) physical world-sheet modes. 
Each term in the above expression is well-known to be divergent. Nevertheless, the whole zero-point energy is of course finite. In appendix \ref{compdelta}, we collect all the formulae required to compute $\Delta$ using zeta-function regularization. The final result is
\be\label{newDelta}
\Delta = 1 - \frac12 \sum_{b=1}^8 \omega_{0b} + \log2 \sum_{b=1}^8 \mu^2_b + \ldots \, , \quad \omega_{0b} = \sqrt{\mu^2_b + \nu^2_b} \, .
\ee
Everything boils down to replacing the masses of the bosonic zero-modes in \eqref{oldDelta} with the corresponding frequencies. Notice that, if the mass- and shift-matching conditions are satisfied, no information about the fermionic sector appears explicitly.

The punchline is that the implicit equation that defines the (inverse) Hagedorn temperature of generic strings is still \eqref{implicitbetaH}, once we use the new expression for the zero-point energy given in \eqref{newDelta} and we take into account how the string tension in \eqref{g000} is modified by the finite chemical potentials (see also \eqref{chargedmapping}). Let us stress that the coefficient in front of the $\log 2$-correction has the geometrical interpretation provided in \eqref{bosonictrace} in terms of the Ricci tensor. This reproduces the very same contribution conjectured in \cite{Harmark:2024ioq}, relying on the principle of covariance in an effective framework. Recently, the same argument has been adopted in \cite{Ekhammar:2025efc}. Here, it comes from a solid world-sheet computation and has a precise physical interpretation. In appendix \ref{app:flatspace}, we successfully apply this method to a rotating type II string in flat space, recovering the result on the Hagedorn temperature (see also \cite{Seitz:2025wpc}) for this case.  

\section{Spinning strings in confining backgrounds}
\label{Sec:confining_bg}

\begin{table}[t]
\centering
\renewcommand{\arraystretch}{1.4}

\begin{tabular}{l!{\vrule width 1.5pt}cc|c!{\vrule width 1.5pt}c}
 & \multicolumn{2}{c|}{Confining backgrounds} & \multirow{2}{*}{Global AdS} & \multirow{2}{*}{Non-contractible} \\ \cline{2-3}
 & \multicolumn{1}{c|}{Minkowski} & Contractible & & \\ \Xhline{1.5pt}
shrinking & S & Q & S & Q \\
non-shrinking & / & / & / & Q
\end{tabular}
\caption{In this table we specify whether the switched-on ``chemical potentials'' couple either to the global $U(1)$ charges (Q) or to the spin (S) of the operators in the dual field theory. Each entry corresponds to a rotating string on a shrinking or not shrinking trajectory in a particular sector of the background geometry. Notice that, in the case of confining backgrounds in the strict sense, the $U(1)$ is understood as a global symmetry from an \emph{infrared} point of view of the (reduced) dual field theory.}
\label{sumtable}
\end{table}

With the aim of applying the previous analysis to relevant classes of examples, in this section we discuss the world-sheet derivation of the Hagedorn temperature of a rotating string in confining backgrounds, including the duals of CFTs on spheres, \ie global $AdS$ spaces.
We consider the string wrapped on the Euclidean temporal circle of length $\beta$ and sitting at the bottom of the space, where the geometry describes the IR regime of the dual field theory.
In that region, the metric typically has some (warped) Minkowski directions, a contractible cycle, and non-contractible cycles. For better clarity, we display in Table~\ref{sumtable} a summary of the different examples analyzed in this section, both for confining backgrounds in the strict sense and global $AdS$.
Finally, 
we consider backgrounds with non-trivial dependence on some non-negative radius, possibly some other internal directions (involved in the rotation) and no Kalb-Ramond field $B$.

In this section we mostly concentrate on the general formulas and provide a few examples.
The results of this section, complemented by the values of $k_2, k_3$, are employed in section \ref{sec:effective} to write the explicit (inverse) Hagedorn temperature in a number of selected cases. 

\subsection{Rotation in Minkowski and in the contractible cycle}
\label{sec:rotMinkDisk}

Let us begin with a string rotating in $n_m$ planes in the $(p+1)$-dimensional Minkowski subspace of the background and in $n_c$ planes in the $d$-dimensional contractible cycle of the geometry.
The configuration is equivalent to a static string on a suitably twisted metric.\footnote{See appendix \ref{app:flatspace} for a physical intuition.}
The asymptotic form of the metric is then taken to be
\begin{subequations}\label{boostedgenericmetric}
\begin{align}
\rmd s^2 = D \, \tilde{g}^{(0)}(y,q) \Biggl[ \rmd & t^2 + \eta_{ij} \rmd x^i \rmd x^j + \sum_{k=1}^{n_m} \Bigl(\rmd\rho_k^2 + \rho_k^2\bigl(\rmd\theta_k -  \omega^{(m)}_k \hspace{-2pt} \rmd t\bigr)^2 \Bigr)\Biggr]\hspace{-2pt} +\\
&+\rmd \vec y^{\, 2} + \sum_{\ell=1}^{n_c} \Bigl(\rmd q_\ell^2 + q_\ell^2 \bigl(\rmd \phi_\ell +i \, \Omega^{(c)}_\ell \hspace{-2pt} \rmd t\bigr)^2\Bigr) + \tilde{g}^{(\mathcal M)}(y,q) \rmd s_{\cal M}^2 \,,\nb
\end{align}
where $i$, $j=1,\dots,p-2\pt n_m$ and 
\begin{align}
&\label{Dandangvel}D \equiv 2\pi \alpha' \pt T_s\,, \quad \beta \pt\omega^{(m)}_k = 2 \pi \nu^{(m)}_k\,, \quad \beta\,\Omega^{(c)}_\ell = 2 \pi \nu^{(c)}_\ell\,,\\[0.5ex]
&\vec y = \{y_1,...,y_{d-2\pt n_c}\} \, , \quad y^2 = \sum_{a=1}^{d-2\pt n_c} y_a^2 \, , \quad q^2 = \sum_{\ell=1}^{n_c} q^2_\ell \, ,\\[0.5ex]
&\tilde{g}^{(0)}(y,q)= 1+\tilde{g}^{(0)}_{2} \bigl(\,y^2+q^2\,\bigr) + \tilde{g}^{(0)}_{4} \bigl(\,y^2+q^2\,\bigr)^2 + \ldots\,, 
\end{align}
\end{subequations}
$\nu^{(m)}_k$ and $\nu^{(c)}_\ell$ being the shifts in the standard bosonic harmonics as in \eqref{genericshiftedharmonics}. Here, we chose to work with imaginary (real) angular velocities $\Omega^{(m)}_k=i\pt\omega^{(m)}_k$ $\bigl(\Omega^{(c)}_\ell\bigr)$ in the Minkowskian (contractible) sector.\footnote{See footnote \ref{imaginaryfoot} for a discussion about it. See also below, under equation \eqref{genimpleqn}.}
Moreover, ${\cal M}$ is a $(9-p-d)$-dimensional non-contractible cycle.
In this scenario, the explicit form of the transverse space $\mathcal M$ does not matter at all (the IR expansion of the  $\tilde{g}^{(\mathcal M)}(y,q)$ factor will be provided later on, in equation \eqref{gM_exp}). At the end of this section, we will discuss some concrete cases that are preparatory for what follows. For the time being, notice that the above metric reduces to the starting point of \cite{Bigazzi:2024biz} (see equation (4.1) there) for $n_m$, $n_c=0$.\footnote{In \cite{Bigazzi:2024biz} it is shown that the results for the Hagedorn temperature can be written in a covariant form. In this paper, for the sake of simplicity, we are not going to present the covariant form of the results.} Moreover, $\eta$ stands for a mostly plus flat Lorentzian metric.\footnote{One of the $x$-directions has been Wick-rotated to preserve the Lorentzian signature of the target space, which is crucial for the Majorana condition on the Green-Schwarz spinors in ten dimensions.}

At the Hagedorn point \eqref{Hagreg}, $\beta=\beta_H$ and the bosonic mass matrix in \eqref{bosmassmatr} takes the form
\footnote{Notice that a minimal string configuration can be taken with an embedding along the $t$ and the Wick-rotated $x$ directions, as, \eg $t= \beta\pt\sigma/2\pi$ and $x= J \pt \tau$ with constant $J$. The Hagedorn limit is taken by sending $\beta\to \beta_H$ and $J \to 0$. Moreover, to have an extremal world-sheet, the string should be placed at $\vec y=0$, $q_l =0$ and at a fixed point in the internal manifold.}
\be\label{rotatingbosmassmat}
\mathcal M_b = \mu^2 \begin{pmatrix}\,\mathbb 0_{p+1} & \, & \, \\ \, & \mathbb 1_d & \, \\ \, & \, & \mathbb 0_{9-p-d}\, \end{pmatrix}\, , \quad \mu^2 = \frac{\beta^2_H}{4\pi^2} D \pt \tilde g^{(0)}_2 \, .
\ee
Here, $\mathbb 0_n$ stands for a $n\times n$ null matrix and $\mathbb 1_d$ denotes the $d\times d$ identity matrix. All the other (omitted) entries are zero. Notice that the masses $\mu$ correspond to the values derived in \cite{Bigazzi:2024biz} in the non-rotating case.\footnote{As it is clear from the example in appendix \ref{app:flatspace}, a simple rotation is expected to leave the world-sheet masses unchanged.} On the other hand, the characteristic polynomial of the connection $A$ defined in \eqref{Hagboseom} is given by\footnote{Notice that using the definitions in \eqref{Hagboseom}, the matrix $A$ might have some components behaving, for instance, as $q_l$ or $1/q_l$, which seems to be pathological for a minimal string at $q_l =0$. A finite connection can be obtained by splitting the fluctuations in longitudinal and transverse modes and projecting $A$ with the normal vectors, as in appendix \ref{app_geom}. In any case, the characteristic polynomial of $A$ will be given as in~\eqref{charpolArotating}.}
\be\label{charpolArotating}
\mathcal P(\lambda)=\lambda^{10-2(n_m+n_c)} \pt \prod_{k=1}^{n_m}\(\lambda^2-\nu^{(m)2}_k\) \prod_{\ell=1}^{n_c}\(\lambda^2+\nu^{(c)2}_\ell\) \, .
\ee
The non-trivial zeros --- related to the directions that span the planes of rotation in pairs --- deform the related harmonics as in \eqref{genericshiftedharmonics}.

The Hagedorn temperature for the rotating string follows immediately from equation \eqref{implicitbetaH}, equipped with the new version of the zero-point energy introduced in \eqref{newDelta}. In more detail, $\beta_H$ is given by the implicit equation
\begin{align}\label{genimpleqn}
\frac{D \pt \beta^2_H}{8\pi^2\alpha'} = 1 - \sum_{k=1}^{n_m} \bigl|\nu^{(m)}_k\bigr| - \Biggl[\,(d-2\pt &n_c)\sqrt{D \pt \tilde g^{(0)}_2}+2\sum_{\ell=1}^{n_c} \sqrt{D \pt \tilde g^{(0)}_2- \Omega^{(c)2}_\ell}\,\Biggr] \frac{\beta_H}{4\pi} +\\
\, &+\( \, d \, \log2 \, D \pt \tilde g^{(0)}_2 + k_2 \,\) \frac{ \beta^2_H}{4\pi^2} + k_3 \frac{ \beta^3_H}{8\pi^3} +\ldots \, , \nb
\end{align}
where $k_2, k_3$ come from the effective approach (see equation (\ref{gendeltaE})).
Notice that some of the eigenvalues of the connection $A$ correspond to imaginary angular velocities. 
Turning the latter to be real might be problematic in some cases \cite{Seitz:2025wpc}: in flat space, for instance, the thermodynamical ensemble at fixed real angular velocity and temperature is unstable. The same instability is instead absent (and the gravitational partition function is well defined) for rotations with subluminal angular velocities along contractible cycles in global $AdS$. 

Notice, moreover, that $\Omega^{(c)}_\ell$ have no reason to be arbitrary small when expanding in powers of $\alpha'$. On the other hand, we can imagine to keep $\nu^{(m)}_k$ fixed in the same perturbative expansion, just like in flat space (cf.~appendix \ref{app:flatspace}). 

All in all, solving the above implicit equation for $T_H=\beta^{-1}_H$, the final result at next-to-next-to-next-to-leading order (NNNLO) in the $\alpha'$ expansion is
\begin{subequations}\label{rotatingTH}
\be\label{expandedTH}
\boxed{
T_H = \frac{\mathfrak t_{\text{LO}}}{\sqrt{\alpha'}} + \mathfrak t_{\text{NLO}} + \mathfrak t_{\text{NNLO}} \sqrt{\alpha'} + \mathfrak t_{\text{NNNLO}} \pt \alpha' + \mathcal O\(\alpha^{3/2}\)
} \, ,
\ee
where
\begin{empheq}[box=\fbox]{align}
&\mathfrak t_{\text{LO}} = \frac{1}{2\pi\sqrt{2}}\sqrt{\frac{D}{1-\sum_{k=1}^{n_m} \bigl|\nu^{(m)}_k\bigr|}} \, , \\[1ex]
&\mathfrak t_{\text{NLO}}=\frac{\pi \pt \mathfrak t_{\text{LO}}^2}{D} \,\Biggl[\,(d-2\pt n_c)\sqrt{D \pt \tilde g^{(0)}_2}+2\sum_{\ell=1}^{n_c} \sqrt{D \pt \tilde g^{(0)}_2- \Omega^{(c)2}_\ell}\,\Biggr]\, ,\\[1ex]
&\label{tNNLOtNNNLO}\mathfrak t_{\text{NNLO}} = \frac{\mathfrak t_{\text{NLO}}^2}{2 \pt \mathfrak t_{\text{LO}}} - \frac{\mathfrak t_{\text{LO}}}{D} \pt k_2 - d \pt \log2 \,\pt \tilde g^{(0)}_2 \, \mathfrak t_{\text{LO}} \,, \quad \mathfrak t_{\text{NNNLO}} = - \frac{k_3}{2\pi\pt D} \, .
\end{empheq}
\end{subequations}
As a first observation, notice that when $\sum_{k=1}^{n_m} \bigl|\nu^{(m)}_k\bigr|\rightarrow 1$ the Hagedorn temperature diverges. In flat space, as discussed in \cite{Seitz:2025wpc, Urbach:2026qph} for the case of a single angular velocity, the reason is that in the limit the partition function computes the supersymmetric index of the theory, losing its divergence at finite $T_H$. We might argue that something similar has to happen in the general case. 

Notice moreover that $\mathfrak t_{\text{NNLO}}$ and $\mathfrak t_{\text{NNNLO}}$ are fixed up to the contributions coming from $\Delta \mathcal E$, which is defined in \eqref{gendeltaE}. On the other hand, the quadratic world-sheet theory is enough to compute both $\mathfrak t_{\text{LO}}$ and $\mathfrak t_{\text{NLO}}$. Also, as already stressed, it turns out to be crucial for finding out the value of the $\log 2$ term in $\mathfrak t_{\text{NNLO}}$, which is missed by the thermal scalar effective approach. The latter will be adopted in section \ref{sec:effective} to compute the values of the $k_2$ and $k_3$ coefficients. However, one could in principle compute them working with a beyond-quadratic order world-sheet theory. 

Now, the time is ripe to check whether our proposal is consistent with some of the results in \cite{Seitz:2025wpc} and \cite{Ekhammar:2025efc}. The authors of \cite{Seitz:2025wpc} focused on generic $d$-dimensional CFTs compactified on $(d-1)$-dimensional spheres, with a single chemical potential coupled to the spins of the operators. There, they found the NLO correction to the Hagedorn temperature in the $\alpha'$-expansion through an effective field theory computation. The authors of \cite{Ekhammar:2025efc} pushed the formalism up to NNNLO for the $d=4$ case, including also another chemical potential in the theory. For $p=n_m=0$, $d=4$, $n_c=2$ and $D=\tilde g^{(0)}_2=1$, our result on $T_H$ reproduces exactly (part of the) formula (3.38) in \cite{Ekhammar:2025efc} (after introducing $g=1/4\pi\alpha'$ and setting the $AdS$ radius $\ell_{AdS}$ to one).\footnote{Notice that the above sigma-model contributions can be safely extrapolated to the limiting regime $\Omega^{(c)2}_\ell \to D\pt \tilde g^{(0)}_2$, for all $\ell$. The breakdown of the perturbative analysis discussed in \cite{Ekhammar:2025efc} within an effective framework affects only the coefficients $k_2$ and $k_3$.} Moreover, the implicit equation in \eqref{genimpleqn} --- for $p=n_m=0$, $D=1$, $\tilde g^{(0)}_2=1/\ell_{AdS}^2$ --- generalizes formula (5.8) in \cite{Seitz:2025wpc} for any eligible value of $n_c$ and gives a prediction for (part of) the NNLO term. 
Finally, notice that the generic LO term in \eqref{genimpleqn} with $D=1$ reproduces the expected result in flat space (cf.~(2.32) of \cite{Seitz:2025wpc}, valid for $n_m=1$; see also \eqref{flatrotatingbetaH}).
\subsection{Rotation in the non-contractible cycles}
\label{sec:WSnoncontr}

So far we dealt with rotations in the contractible cycle of the geometry. It is time to also consider a proper motion of the string in the non-contractible cycle $\mathcal M$. The latter can vary dramatically depending on the specific example. The only universal aspect of this sector lies in the overall factor that can be expanded as\footnote{Note that the radius $R_{\cal M}$ of the manifold ${\cal M}$ was taken to be one in \cite{Bigazzi:2024biz}, since the results of that paper do not depend on its value. In the present context, where we are going to consider rotations in this manifold, the radius will be relevant, so we keep it explicit.}
\be
\label{gM_exp}
\tilde g^{(\cal M)}(y,q) = R^2_{\cal M} \, \Bigl[1+\tilde{g}^{(\cal M)}_{2} \bigl(\,y^2+q^2\,\bigr)^2 + \ldots\Bigr] \, ,
\ee
and in the Hagedorn values of the parameters
\be\label{generalmasses}
\mu^2_{\cal D} = \frac{\beta^2_H}{4\pi^2} \(D \pt \tilde g^{(0)}_2 - R^2_{\cal M} \pt \tilde g^{(\cal M)}_2  \kappa^2 \) \, , \quad \mu^2_{\cal M} = \frac{\beta^2_H \pt \kappa^2}{4\pi^2} \,,
\ee
where $\kappa$ is a certain chemical potential. The latter will be respectively related to the masses of the bosonic fluctuations along the contractible and the non-contractible cycle in the background. We will now proceed by examining the relevant examples in the literature on a case-by-case basis.

\subsubsection{Example: $U(1)$-charge twist in spherical sectors
}
\label{sec:spinonSn}

Let us focus on transverse sectors given by $(9-p-d)$-dimensional spheres, namely $\mathcal M = S^{9-p-d}$, for $9-p-d\ge2$. 

The metric of a unit $n$-dimensional sphere can be recursively written as
\be\label{polarsphere}
\rmd \Omega^2_n =\rmd \theta^2 + \sin^2 \hspace{-1pt}\theta \, \rmd \Omega^2_{n-1} \, ,
\ee
for some polar angle $\theta \in [0,\pi]$. The very same metric can be rephrased in a conformally flat fashion by switching to the stereographic coordinates
\be\label{stereocoord}
\varphi= \tan \theta/2 \in [0,\infty) \, , \quad \varphi^2 = \sum_{j=1}^n \left(\varphi^j\right)^2 \, ,
\ee
which produce
\be\label{stereometric}
\rmd \Omega^2_n =\frac{4}{(1+\varphi^2)^2} \[\rmd \varphi^2 + \varphi^2 \rmd \Omega^2_{n-1}\] = \frac{4 \, \rmd \vec \varphi^{\hspace{1pt}2}}{(1+\varphi^2)^2} \, .
\ee
Another useful coordinate system relies on the recursive relation in \eqref{polarsphere} and gives
\be\label{toroidalsphere}
\rmd \Omega^2_n =\rmd \theta^2 + \sin^2 \hspace{-1pt} \theta \(\rmd \tilde\theta^2 + \sin^2 \hspace{-1pt} \tilde\theta \rmd \Omega^2_{n-2}\) = \sin^2 \hspace{-1pt} \phi \, \rmd \psi^2 + \rmd \phi^2 + \cos^2 \hspace{-1pt} \phi \, \rmd \Omega^2_{n-2} \, ,
\ee
for\footnote{We adopted the two-argument arctangent notation of Wolfram Mathematica.}
\be\label{toroidalcoordinate}
\theta=\arccos (\sin \phi \cos \psi) \, , \quad \tilde\theta = \arctan (\sin \phi \sin \psi, \cos\phi) \, .
\ee
Notice that any slice at fixed $\phi$ has the topology of a generalized torus $S^1 \times S^{n-2}$. For this reason, we will refer to the variables $\phi$, $\psi$ as \emph{toroidal coordinates}. Let us stress that $\phi \in \[ 0, \pi/2\]$ and $\psi \in [0,2\pi)$.

To begin with, we can consider a string rotating in $n_S$ planes in the sphere. The stereographic coordinates introduced above turn out to be very useful for this purpose. Indeed, the twisted metric of the sphere can be written as \footnote{Here, the classical solution is understood as set at the origin of the stereographic coordinates, namely $\varphi^2 = \eta^2 =0$.  Then all the $\chi_a$ are associated with contractible cycles.}
\begin{subequations}\label{boostedsphere}
    \be
\rmd s^2_{\mathcal M} = \frac{4}{\left(1 + \varphi^2+\eta^2\right)^2} \left [ \rmd \vec \varphi^{\, 2} + \sum_{a=1}^{n_S} \Bigl(\rmd \eta_a^2 + \eta_a^2 \(\rmd \chi_a - \omega^{(S)}_a \rmd t\)^2\Bigr) \right ] \hspace{-2pt},
\ee
where\be
\vec \varphi = \{\varphi_1,\ldots,\varphi_{9-p-d-2\pt n_S}\}\, , \quad \varphi^2 = \sum_{m=1}^{9-p-d-2\pt n_S} \varphi_m^2 \, , \quad \eta^2 = \sum_{a=1}^{n_S} \eta^2_a \, .
\ee
\end{subequations}
Then, in this case, one should think of a string that rotates within two-dimensional conformally flat submanifolds of the sphere.

The bosonic mass matrix in the Hagedorn regime looks exactly as in \eqref{rotatingbosmassmat}. On the other hand, the characteristic polynomial of the connection $A$ in \eqref{charpolArotating} must be adjusted as
\be
\mathcal P(\lambda) \mapsto \mathcal P (\lambda) \times \lambda^{-2 \pt n_S} \pt \prod_{a=1}^{n_S}\Bigl(\lambda^2-\nu^{(S)2}_a\Bigr) \, ,
\ee 
where $2\pi\nu^{(S)}_a=\beta_H\pt\omega^{(S)}_a$. Therefore, we conclude that \eqref{rotatingTH} is still valid as long as we shift $\mathfrak t_{\text{NLO}}$ as
\be\label{tNLOstereosphere}
\mathfrak t_{\text{NLO}}\mapsto\mathfrak t_{\text{NLO}}+\frac{2\pi\pt\mathfrak t_{\text{LO}}^2}{D} \sum_{a=1}^{n_S} \bigl|\omega^{(S)}_a\bigr| \, .
\ee
Crucially, the $\log2$-terms are not affected since the modes on the spheres remain massless. 
This result applies, \eg to rotations in Witten's holographic Yang-Mills model (WYM) \cite{Witten:1998zw}. In this case \cite{Bigazzi:2024biz} 
\be\label{WYMparameters}
d=2\,,\quad p=3\,,\quad D =\frac{8 }{27}R^3 M_{KK}^3 \, , \quad    \tilde{g}^{(0)}_{2}= \frac{27}{16 M_{KK} R^3}\,, \quad \tilde{g}^{(\cal M)}_{2} =\frac{9}{16 M_{KK} R^3}\,,
\ee
where $M_{KK}$ denotes the Kaluza-Klein (and glueball) mass scale, while $R$ is the characteristic radius of the spacetime.

The twist in \eqref{boostedsphere} is not the only operation we can apply to the metric of a sphere. Let us restart the analysis in the toroidal coordinates introduced in \eqref{toroidalsphere}. For instance, we can perform the following transformation
\be\label{chargedshift} 
\psi \mapsto \psi + i \pt \kappa \, t \, .
\ee
This translates into a motion of the string along the stereographic radius of the sphere (see appendix \ref{app:Rcharged}). The chemical potential $\kappa$ will be conjugate to some global $U(1)$ charge in the dual QFT. In supersymmetric contexts the global $U(1)$ symmetry may correspond to an abelian subgroup of the R-symmetry group. 

This time, the bosonic mass matrix in the Hagedorn limit\footnote{In this specific example, classically, $\phi$ is non-zero. The solution $\phi=\pi/2$ puts the classical string at the origin of the stereographic coordinate system \eqref{stereometric}, as follows from \eqref{stereocoord} and \eqref{toroidalcoordinate}.} is modified by the presence of the chemical potential. In general, it turns out to be non-diagonal with a characteristic polynomial given by
\be\label{chargedspectrum}
\mathcal Q (\lambda) = \lambda^{p+2} \(\lambda-\mu^2_{\cal D}\)^d \(\lambda- \mu^2_{\cal M}\)^{8-p-d} \, ,
\ee
where the values of $\mu_{\mathcal D}$ and $\mu_{\mathcal M}$ are given in \eqref{generalmasses}. Notice that we recover the scenario in \eqref{rotatingbosmassmat} whenever $\kappa$ is set to zero.
On the other hand, the characteristic polynomial of $A$ in \eqref{charpolArotating} is unaffected.\footnote{Here we are having in mind a string embedding along $t= \frac{\beta}{2\pi}\sigma$ and $\psi =J\tau$. Also in this case, in the Hagedorn limit we set $\beta\to \beta_H$ and $J\to 0$.} 

All in all, the final result for the Hagedorn temperature can be derived from \eqref{rotatingTH} by first shifting $\mathfrak t_{\text{NLO}}$ and $\mathfrak t_{\text{NNLO}}$ as
\begin{subequations}\label{chargedshifting}
    \begin{align}
        &\mathfrak t_{\text{NLO}}\mapsto\mathfrak t_{\text{NLO}}+\frac{\pi \pt \mathfrak t_{\text{LO}}^2}{D} (8-p-d) \pt |\kappa| \, ,\\
        &\mathfrak t_{\text{NNLO}}\mapsto\mathfrak t_{\text{NNLO}}- \frac{\mathfrak t_{\text{LO}}}{D} \pt(8-p-d) \, \kappa^2 \log2 \, ,
    \end{align}
\end{subequations}
and then, in the resulting formulae, by making the substitution
\be\label{chargedmapping}
D \mapsto D\(1-\frac{R^2_{\cal M} \kappa^2}{D}\) \, , \quad \tilde g^{(0)}_2 \mapsto  \(1-\frac{R^2_{\cal M} \kappa^2}{D}\)^{-1} \(\tilde g^{(0)}_2 - \frac{R^2_{\cal M} \tilde g^{(\cal M)}_2 \kappa^2}{D} \) \, .
\ee
Notice that the same result holds even if $\cal M$ is a one-dimensional cycle (for $p+d=8$), with
\be\label{boostedS1metric}
\rmd s^2_{\cal M} =  (\rmd\psi + i \pt \kappa \rmd t)^2 \, .
\ee
In this case there are no extra massive modes and indeed the shifts in \eqref{chargedshifting} cancel out. On the other hand, the map \eqref{chargedmapping} modifies the outcome for $T_H$ in a non-trivial way.

As an example, let us set
\begin{subequations}\label{AdS5xS5parameters}
    \begin{align}
        &d=4 \, , \quad p=n_m=n_c=0 \, , \quad D=1 \, ,\\
        &\tilde g^{(0)}_2 = 1/\ell^2_{AdS} \, , \quad R_{\cal M}= \ell_{AdS} \, , \quad g^{(\cal M)}_2 = 0 \, ,
    \end{align}
\end{subequations}
so as to work with the global-$AdS_5\times S^5$ solution. The dual CFT compactified on $S^3$ features a non-zero chemical potential $\kappa$ that couples to a $U(1)$ subgroup of the total $SO(6)$ R-symmetry group. The final result for the Hagedorn temperature is
\be\label{AdS5chargedTH}
T_H = \frac{\sqrt{1-\bar \kappa^2}}{2\pi\sqrt{2\hspace{1pt}\alpha'}} + \frac{1+|\bar \kappa|}{2\pi\ell_{AdS}} + \frac{(1+|\bar \kappa|)^2 - (1+\bar \kappa^2) \,4\log2 }{2\pi\sqrt{2(1-\bar \kappa^2)}} \frac{\sqrt{\alpha'}}{\ell^2_{AdS}} + \mathcal O \(\alpha'\) \, ,
\ee
where we defined $\bar \kappa = \ell_{AdS} \pt  \kappa$ in order to declutter the equation. It reproduces exactly formula (3.16) in \cite{Ekhammar:2025efc} up to NNLO in the $\alpha'$ expansion, $k_2$ being zero (as a consequence of formula (\ref{effk2k3})).\footnote{The fact that $k_2=0$ in this case should be a consequence of the  extremely simple structure of the world-sheet spectrum. From an effective point of view, the quartic contribution that fix $\Delta \mathcal E$ arises as the vacuum expectation value of the operator in (3.5) of \cite{Ekhammar:2025efc}. At first glance, it should vanish given the expressions in (3.6) there.} 

To wrap up, we saw that different twists on a sphere may correspond to different effects on the world-sheet spectrum. The latter can be combined starting from a clever parameterization of the untwisted metric. To fix ideas, let us work again with the example in \eqref{AdS5xS5parameters}. The round metric on the unit $S^5$ can be written as (see, \eg \cite{Cvetic:1999xp, Chatzis:2024top, Chatzis:2024kdu, Kumar:2024pcz, Castellani:2024ial})
\be\label{metricS5inIR6}
\rmd s^2_{S^5} = \sum_{i=1}^3 \(\rmd \mu_i^2 + \mu^2_i \, \rmd \phi^2_i\) \, , \quad \sum_{i=1}^3 \mu^2_i = 1 \, ,
\ee
where $\phi_1$, $\phi_2$, $\phi_3 \in [0,2\pi)$. For instance, we can choose
\be\label{mu1mu2mu3}
\mu_1 = \sin\theta \sin\varphi \, , \quad \mu_2 = \sin\theta \cos\varphi \, , \quad \mu_3 = \cos\theta \, , \quad \theta , \, \varphi \in [0,\pi/2] \, .
\ee
The above metric features a manifest $U(1)\times U(1)\times U(1)$ isometry, which corresponds to the maximal torus of the $SO(6)$ R-symmetry group. From a string perspective, the $U(1)\times U(1)\times U(1)$ group is associated with rotations around the three independent axes of $S^5$ embedded in $\mathds R^6$.

The metric in \eqref{metricS5inIR6} can be twisted with time as
\be\label{U13shifts}
\phi_i \mapsto \phi_i - \omega_i \, t \, , \quad i=1, 2, 3 .
\ee
Here, $\omega_i$, $i=1$, $2$, $3$, are associated with the (imaginary) angular velocities of the string on the orthogonal planes of rotation within $S^5$ according to the discussion in footnote \ref{imaginaryfoot}. In this scenario, both the characteristic polynomial of $\mathcal M_b$ and that of $A$ are modified by such a transformation. In more detail, defining
\be
2\pi\nu_i = \beta_H \pt \omega_i \, , \quad i=1,2,3\,,
\ee 
they respectively read\footnote{The string can be placed in either the $\theta=0$, $\varphi=0$ or $\theta=0$, $\varphi=\pi/2$. 
Note that with this parametrization we are not treating the three $U(1)$s on the same footing, so the effects of the three chemical potentials on the Hagedorn temperature is going to be different. 
Moreover, given the parameters in \eqref{AdS5xS5parameters}, the masses $\mu_{\mathcal D}$ reduces to $\mu_{AdS} = \beta_H/(2\pi\ell_{AdS})$.}
\begin{subequations}\label{fullychargedspectrum}
    \begin{align}
        &\mathcal Q(\lambda) = \lambda^2 \(\lambda-\mu^2_{\cal D}\)^4 \(\lambda + \nu_3^2\)^4 \, ,\\[1ex]
        &\mathcal P (\lambda) = \lambda^6 \(\lambda^2 - \nu_1^2\) \(\lambda^2 - \nu_2^2\) \, .
    \end{align}
\end{subequations}
The extra massive modes and the non-trivial shifts in the harmonics both refer to the directions of the sphere, except for $\phi_3$. We conclude that $\omega_1$ ($\omega_2$) plays the same role of the angular velocity $\omega_1^{(s)}$ ($\omega_2^{(s)}$) in \eqref{boostedsphere}, while $\omega_3$ mimics the chemical potential $\kappa$ in \eqref{chargedshift}. This is even more clear looking at the embedding maps of the sphere shown in appendix \ref{app:Rcharged}.

The Hagedorn temperature in such a generic scenario reads
\begin{align}\label{finalchargedTH}
    T_H = &\frac{\sqrt{1+\bar \omega_3^2}}{2\pi\sqrt{2\hspace{1pt}\alpha'}} + \frac{1+\frac12\sqrt{\bar\omega_1^2-\bar\omega_3^2}+\frac12\sqrt{\bar\omega_2^2-\bar\omega_3^2}}{2\pi\ell_{AdS}} + \\
    &+\frac{\Bigl(1+\frac12\sqrt{\bar\omega_1^2-\bar\omega_3^2}+\frac12\sqrt{\bar\omega_2^2-\bar\omega_3^2}\Bigr)^2 \hspace{-4pt}- k_2 \pt \ell_{AdS}^2 - (1-\bar \omega_3^2) \,4\log2}{2\pi\sqrt{2(1+\bar \omega_3^2)}} \frac{\sqrt{\alpha'}}{\ell^2_{AdS}} + \mathcal O \(\alpha'\),\nonumber
\end{align}
where, as before, we simplified the outcome by introducing $\bar\omega_i=\ell_{AdS} \, \omega_i$, $i=1$, $2$, $3$. It readily follows that the result produced by the shift \eqref{tNLOstereosphere} in $\mathfrak t_{\text{NLO}}$ corresponds to the above expression once we set
\be\label{fromgentostereo5}
\omega_1=\omega_1^{(S)} \, , \quad \omega_2=\omega_2^{(S)} \, , \quad \omega_3 = 0 \, .
\ee
On the other hand, we recover the final result in \eqref{AdS5chargedTH} if
\be\label{fromgentotoro5}
\omega_1=\omega_2=0 \, , \quad \omega_3 \mapsto \pm i \, \kappa\,.
\ee
We conclude that \eqref{finalchargedTH} corresponds to the Hagedorn temperature of $\mathcal N=4$ Super Yang Mills at finite temperature with chemical potentials that couple to any combination of the $U(1)$ subgroups of $SO(6)$.

Of course, the above can be generalized to any sphere of odd dimension $n$ with whatever warp factor. In this case, the metric of the internal space reads
\be\label{metricSninIRnpiu1}
\rmd s^2_{S^n} = \sum_{i=1}^{(n+1)/2} \bigl(\rmd \mu_i^2 + \mu^2_i \, \rmd \phi^2_i\bigr) \, , \quad \sum_{i=1}^{(n+1)/2} \mu^2_i = 1 \, .
\ee
Therefore, its twisted version is nothing but
\begin{align}
    \rmd s^2_{S^n} = &\sum_{i=1}^{(n-1)/2}  \Bigl[ \bigl(1+4\pt\mu^2_i\bigr)\rmd \mu_i^2 + \mu^2_i \, \bigl(\rmd \phi_i - \omega_i \pt \rmd t\bigr)^2\Bigr] + \\[1ex]
    &+ 8 \sum_{\substack{i,j=1 \\ i > j}}^{(n-1)/2 } \mu_i \pt \mu_j \rmd \mu_i \rmd \mu_j +\Biggl( 1 - \sum_{i=1}^{(n-1)/2} \mu_i^2 \Biggr) \bigl(\rmd \phi_{(n+1)/2} - \omega_{(n+1)/2} \pt \rmd t\bigr)^2 \, . \nonumber
\end{align}
Notice that we have already implemented the constraint in \eqref{metricSninIRnpiu1}.

For generic Minkowski and contractible sectors, it follows that the characteristic polynomial of the bosonic mass matrix is\footnote{The classical string configuration features $\mu_i=0$, $\forall$ $i=1,\ldots,(n-1)/2$, and $\mu_{(n+1)/2}=1$.}
\be
\mathcal Q (\lambda) = \lambda^{p+2} \(\lambda-\tilde\mu^2_{\cal D}\)^d \(\lambda- \tilde\mu^2_{\cal M}\)^{n-1} \, ,
\ee
where $\tilde\mu_{\mathcal D}$ and $\tilde\mu_{\mathcal M}$ are respectively given by $\mu_{\mathcal D}$ and $\mu_{\mathcal M}$ in \eqref{generalmasses} with
\be
\omega_{(n+1)/2}=\pm i\pt\kappa \, .
\ee
Let us stress that the only massless boson on the sphere is related to the $\phi_{(n+1)/2}$-direction. On the other hand, the characteristic polynomial of the connection $A$ gets modified as
\be
\mathcal P(\lambda) \mapsto \mathcal P (\lambda) \times \lambda^{-(n-1)} \pt \prod_{i=1}^{(n-1)/2}\Bigl(\lambda^2-\nu^2_i\Bigr) \, .
\ee
So, all the bosonic harmonics associated with the directions of the sphere other than $\phi_{(n+1)/2}$ are shifted.

As before, we can compute the Hagedorn temperature in this very general framework starting from \eqref{rotatingTH}. First, $\mathfrak t_{\text{NLO}}$ and $\mathfrak t_{\text{NNLO}}$ must be shifted as
\begin{subequations}
    \begin{align}
        &\mathfrak t_{\text{NLO}}\mapsto\mathfrak t_{\text{NLO}}+\frac{2\pi \pt \mathfrak t_{\text{LO}}^2}{D} \sum_{i=1}^{(n-1)/2} \sqrt{\omega_i^2-\omega_{(n+1)/2}^2} \, ,\\
        &\mathfrak t_{\text{NNLO}}\mapsto\mathfrak t_{\text{NNLO}} + \frac{\mathfrak t_{\text{LO}}}{D} \pt (n-1) \, \omega_{(n+1)/2}^2 \log2 \, ,
    \end{align}
\end{subequations}
Then, in the resulting formulae, we need to perform the following substitutions
\be
D \mapsto D\(1+\frac{R^2_{\cal M} \pt \omega_{(n+1)/2}^2}{D}\) , \quad \tilde g^{(0)}_2 \mapsto  \(1+\frac{R^2_{\cal M} \pt \omega_{(n+1)/2}^2}{D}\)^{-1} \hspace{-2pt} \(\tilde g^{(0)}_2 + \frac{R^2_{\cal M} \tilde g^{(\cal M)}_2 \pt \omega_{(n+1)/2}^2}{D} \) .
\ee

These prescriptions both include \eqref{tNLOstereosphere} for
\be
\omega_i=\omega_i^{(S)} \, , \quad \forall i=1,\ldots,n_S=(n-1)/2 \, , \quad \omega_{(n+1)/2}=0 \, ,
\ee
and \eqref{chargedshifting}-\eqref{chargedmapping} for 
\be
\omega_i=0 \, , \quad \forall i=1,\ldots,n_S=(n-1)/2 \, , \quad \omega_{(n+1)/2}\mapsto\pm i \pt \kappa \, ,
\ee
generalizing the relation in \eqref{fromgentostereo5} and \eqref{fromgentotoro5} for $n=5$. This formalism allowed us to work with the maximum possible number of planes of rotation in a sphere of odd dimension. Notice that, in the case of an even dimension, we already had access to this number in stereographic coordinates (see \eqref{boostedsphere}).

\subsubsection{Example: rotation along the Sasaki-Einstein Reeb vector}
\label{sec:reeb}
Let $\kappa$ be the chemical potential for the $U(1)_R$ R-symmetry of CFTs (on spheres) dual to $AdS_5 \times X^5$, where $\mathcal M = X^5$ is a five-dimensional Sasaki-Einstein (SE) manifold.\footnote{Further examples with different dimensionalities might be considered too, starting for instance from the results in \cite{Cvetic:2005ft}.} 
Our formalism can be used to obtain the dependence of the Hagedorn temperature on $\kappa$ in these scenarios. Indeed, the $U(1)_{\text R}$ is dual to an isometry generated by the Reeb vector $\partial_\psi$ and the generic (twisted) metric can be written as
\be\label{SEmetric}
\rmd s^2_{X^5} = \left(\rmd\psi + i \, \mathcal \kappa \rmd t + a \right)^2 + \rmd s^2_{\mathcal N}\,,
\ee
where $a$ is a certain one-form depending on the coordinates of the four-dimensional K\"ahler-Einstein manifold $\mathcal N$, with K\"ahler two-form proportional to $\rmd a$. Technically, we can say that $\mathcal M$ is a (twisted) $U(1)_{\text{R}}$ bundle over $\mathcal N$.

The infinite family of examples to which this applies includes the toric SE spaces $L^{p,q,r}$~\cite{Cvetic:2005ft}, where $p$, $q$, $r$ are coprime positive integers such that $0 < p \leq q$, $0 < r < p+q$. In this case, the metric of $\mathcal N$ and the one-form $a$ read\footnote{One can get rid of all the trigonometric functions introducing $z=\alpha \cos^2\theta + \beta \sin^2\theta$. This is a very helpful change of coordinates for calculating the geometric properties of the background.}
\begin{subequations}
\begin{align}
    &\rmd s^2_{\mathcal N} = \frac{\rho^2 \rmd x^2}{4 \pt \Delta_x} + \frac{\rho^2\rmd \theta^2}{\Delta_\theta} + \frac{\Delta_x}{\rho^2} \Biggl(\frac{\sin^2\theta}{\alpha}\rmd\phi_1 + \frac{\cos^2\theta}{\beta}\rmd\phi_2\Biggl)^{\hspace{-3pt}2}+\\
    &\qquad\qquad\qquad\quad + \frac{\Delta_\theta \sin^2\theta\cos^2\theta}{\rho^2} \Biggl(\frac{\alpha-x}{\alpha}\rmd\phi_1 + \frac{\beta-x}{\beta}\rmd\phi_2\Biggr)^{\hspace{-3pt}2} \, ,\nonumber\\[1ex]
    &a=\frac{(\alpha-x)\sin^2\theta}{\alpha} \rmd\phi_1 + \frac{(\beta-x)\cos^2\theta}{\beta} \rmd\phi_2\, ,
\end{align}
where
\begin{align}
&\Delta_x = x (\alpha-x) (\beta - x) -\gamma \, , \quad \Delta_\theta = \alpha \cos^2\theta + \beta \sin^2\theta \, , \\[1ex]
&\rho^2 = \Delta_\theta - x \, , \quad 0<\theta<\pi/2 \, , \quad x_1 < x < x_2 \, ,
\end{align}
\end{subequations}
$x_1$ and $x_2$ being the smallest roots of $\Delta_x$. Notice that we have the freedom to set one of the parameters $\alpha$, $\beta$, $\gamma$ to any non-zero constant, just by rescaling the other two and $x$. Therefore, the above metrics depend on just two non-trivial parameters. The latter, together with $x_1$ and $x_2$, can be expressed as functions of the triple $(p,q,r)$ thanks to the completeness and non-singularity constraints on the manifold. 

The characteristic polynomials of the bosonic mass matrix and of the connection $A$ can be computed even on this very general background, with parameters \eqref{AdS5xS5parameters} for the $AdS$ sector. 
In the Hagedorn regime, they respectively read\footnote{Here, the Hagedorn regime is obtained by taking the limit $J\to 0$, where $J$ is the momentum associated with the Reeb vector.}\footnote{As in \eqref{fullychargedspectrum}, on the parameters in \eqref{AdS5xS5parameters}, the masses $\mu_{\mathcal D}$ corresponds to $\mu_{AdS} = \beta_H/(2\pi\ell_{AdS})$.}
\begin{subequations}\label{SEspectrum}
    \be
\mathcal Q (\lambda) = \lambda^2 \(\lambda-\mu^2_{\mathcal D}\)^4 \(\lambda - \mu^2_{\mathcal M}\)^4
\ee
and
\be
\mathcal P(\lambda)=\lambda^6 \(\lambda^2+\nu^2_{\mathcal M}\)^2\, , \quad \nu^2_{\mathcal M} = - \mu^2_{\mathcal M} \, . 
\ee
\end{subequations}
Notice that the non-trivial eigenvalues of $\mathcal P$ are related to the massive modes on $\mathcal M$. Therefore, in this case, the above negative shifts squared cancel each other out with the corresponding world-sheet masses in the zero mode frequencies \eqref{newDelta}. We conclude that the Hagedorn temperature is universally given by
\be\label{reebTH}
T_H = \frac{\sqrt{1-\bar \kappa^2}}{2\pi\sqrt{2\hspace{1pt}\alpha'}} + \frac{1}{2\pi\ell_{AdS}} + \frac{1 - k_2 \pt \ell^2_{AdS} - (1+\bar \kappa^2) \,4\log2 }{2\pi\sqrt{2(1-\bar \kappa^2)}} \frac{\sqrt{\alpha'}}{\ell^2_{AdS}} 
-\frac{k_3}{2\pi (1-\bar\kappa^2)}\alpha'\,,
\ee
where (see section \ref{sec:effective} for the results about $k_2, k_3$) 
\be 
\bar \kappa = \ell_{AdS} \pt  \kappa\,, \qquad k_2= -\frac32 (1-\bar \kappa^2)\,, \qquad k_3 = -\frac{3}{16}(1-\bar \kappa^2)(15+\bar \kappa^2)\,.
\ee 
This result does not (explicitly) depend on $\alpha$, $\beta$ and $\gamma$. Therefore, at least formally, it holds universally for any triple $(p,q,r)$. However, each case corresponds to a different value of $\ell_{AdS}$ that relies on the volume of the internal space (see e.g.~\cite{Gubser:1998vd}).

The most notable examples of SE manifolds are the base of the conifold
$T^{1,1}$ and the five-sphere $S^5$ (\eg see \cite{Gauntlett:2004yd, Cvetic:2005ft, Butti:2005sw}). Indeed, for $\alpha=\beta$, we get
\be\label{T11metric}
\rmd s^2_{T^{1,1}} = \frac19\left(\rmd\psi + 3 \pt i \kappa \rmd t + \cos{\theta_1} \rmd\phi_1 + \cos{\theta_2} \rmd\phi_2  \right)^2 + \frac16 \rmd\Omega_{2,1}^2+ \frac16 \rmd\Omega_{2,2}^2 \,,
\ee
which is the homogeneous metric on the  $L^{1,1,1}=T^{1,1}$ space with $S^2\times S^2$ as the base. Indeed, $\Omega_{2,1}$ and $\Omega_{2,2}$ are round metrics on unit two-spheres with coordinates $\{\theta_1,\phi_1\}$ and $\{\theta_2, \phi_2\}$ respectively, while $\psi$ is an angular variable of period $4\pi$. On the other hand, in the limit $\gamma=0$, the round metric on $L^{0,2,1}=S^5$ can be expressed as in \eqref{SEmetric} with the four-dimensional base space $\mathcal N$ being $\mathds C \text P^2$, namely
\begin{align}\label{SEmetricS5}
    \rmd s^2_{S^5} =  &\left[\rmd\psi + i \pt \kappa \rmd t -\frac12 \sin^2\chi \(\rmd \alpha_3 + \cos\alpha_1 \rmd \alpha_2 \) \right]^2 + \rmd \chi^2 + \\
    &+\frac14 \sin^2\chi \(\rmd\alpha_1^2+\sin^2\alpha_1\rmd\alpha_2^2\)+ \frac14 \sin^2\chi \cos^2\chi \(\rmd\alpha_3+\cos\alpha_1\rmd\alpha_2\)^2 \, , \nonumber
\end{align}
where the period of $\psi$ and $\alpha_3$ is taken to be $2\pi$ and $4\pi$ respectively. Notice that the general result for the Hagedorn temperature of this section differs from what we get in \eqref{AdS5chargedTH}, already at NLO. The reason is that the $U(1)_R$ symmetry subgroup corresponding to the Sasaki-Einstein Reeb vector does not coincide with the R-symmetry subgroup $U(1)\in SO(6)$ considered there and in \cite{Ekhammar:2025efc}. 
Under the former, in fact, the three complex scalars of ${\cal N}=4$ SYM, seen, through the dual SE parameterization, as a ${\cal N}=1$ theory with cubic superpotential, have charge 2/3. Under the latter, instead, only one complex scalar is charged, with unit charge, while the remaining two are uncharged. Consistently, the result in \eqref{reebTH} can be derived from the general proposal in \eqref{finalchargedTH} by setting there
\be
\omega_1=\omega_2=\omega_3\mapsto \pm i\pt\kappa\,.
\ee
This is in agreement with the discussion in appendix \ref{app:Rcharged} (cf.~\eqref{embedS5SE} with \eqref{toroidalembed}).


\subsubsection{Example: rotation in complex projective spaces}
\label{sec:WSCpn}

The twisted metric on the unit complex projective space $\mathcal M = \CP^n$ can be written iteratively as (see e.g.~\cite{Hoxha:2000jf})
\be
\rmd s^2_{\mathds C \text P^n} = \rmd \xi^2 + \sin^2\xi \cos^2 \xi \(\rmd\psi+2\pt i \, \mathcal \kappa \rmd t + a \pt\)^2 + \sin^2\xi \rmd s^2_{\mathds C \text P^{n-1}} \, ,
\ee
for $\xi\in[0,\pi/2]$, where
\be
\rmd s^2_{\CP^{n-1}} = f^{-2} \sum_{i=1}^{n-1} \Bigl(f \pt \rmd v^*_i \rmd v_i - |v^*_i \rmd v_i|^2\Bigr)  \, , \quad f = 1+\sum_{i=1}^{n-1} v^*_i v_i \, ,
\ee
is the Fubini-Study metric on the unit $\CP^{n-1}$ and
\be
a=\frac{i}{2\pt f} \, \sum_{i=1}^{n-1} \bigl(\,v_i \rmd v^*_i - v^*_i \rmd v_i\,\bigr) 
\ee
is a potential for the K\"ahler form of $\CP^{n-1}$. Notice that $v_i$ are complex variables, $v^*_i$ being the corresponding complex conjugates. In principle, one should switch to real coordinates $\{r_i,\theta_i\}_{i=1}^{n-1}$ with $v_i=r_i e^{i\pt\theta_i}$. The result is
\be
f^2 \pt \rmd s^2_{\CP^{n-1}} = \sum_{i=1}^{n-1} \(f-r^2_i\) \(\rmd r_i^2 + r_i^2 \rmd \theta_i^2\) - 2 \sum_{\substack{i,j=1 \\ i > j}}^{n-1
} r_i \pt r_j \(\rmd r_i \rmd r_j + r_i r_j \rmd \theta_i \rmd \theta_j \) \, ,
\ee
with $f=1+\sum_i^{n-1} r^2_i$. This is a very useful expression for calculations, along with the substitution $z = \cos\xi$.

It is possible to compute the characteristic polynomial of the bosonic mass matrix for any given $n$. Let us stress that we are interested in the case in which $9-p-d=2n$ is even. We get\footnote{Classically, the string configuration features a non-zero value for the direction $z=\cos\xi$, \ie $z=1/\sqrt{2}$.}
\be
\mathcal Q (\lambda) = \lambda^{p+2} \(\lambda-\mu^2_{\cal D}\)^d \(\lambda-4\,\mu^2_{\mathcal M}\) \(\lambda- \mu^2_{\mathcal M}\)^{2(n-1)} \, .
\ee
Here, $2\pt\mu_{\mathcal M}$ represents the mass of the bosonic fluctuation along $z=\cos\xi$, while the other $2(n-1)$ masses $\mu_{\mathcal M}$ are related to the directions of $\CP^{n-1}$. Similarly to the previous example, the characteristic polynomial of the connection $A$ in \eqref{charpolArotating} is deformed as
\be
\mathcal P(\lambda) \mapsto \mathcal P(\lambda) \times \lambda^{-2(n-1)} \pt \Bigl(\lambda^2+\nu^2_{\mathcal M}\Bigr)^{\hspace{-2pt}n-1} \, , \quad \nu^2_{\mathcal M} = - \mu^2_{\mathcal M} \, .
\ee
The new non-trivial eigenvalues all refer to the $\CP^{n-1}$ sector of the geometry. Again, some of the frequencies of the zero modes in \eqref{newDelta} cancel out.

Therefore, the Hagedorn temperature readily follows from \eqref{rotatingTH} by first shifting $\mathfrak t_{\text{NLO}}$ and $\mathfrak t_{\text{NNLO}}$ as
\begin{subequations}
    \begin{align}
        &\mathfrak t_{\text{NLO}}\mapsto\mathfrak t_{\text{NLO}}+\frac{\pi \pt \mathfrak t_{\text{LO}}^2}{D} \, 2 \, |\kappa| \, ,\\
        &\mathfrak t_{\text{NNLO}}\mapsto\mathfrak t_{\text{NNLO}}- \frac{\mathfrak t_{\text{LO}}}{D} \, 2 \, (n+1) \, \kappa^2 \log2 \, ,
    \end{align}
\end{subequations}
and then, in the resulting formulae, by making the very same substitution as in \eqref{chargedmapping}. 
This result can be applied to the ABJM case by setting $p=0$, $d=3$, $n=3$ and constant $\tilde g^{(\mathcal M)}=R^2_{\mathcal M}=4\, \ell_{AdS}^2$~\cite{Aharony:2008ug}.

\subsubsection{Example: rotations in a torus}

We can think of introducing a toric sector into the background via the replacement
\be
\mathcal M \mapsto \mathcal M \times T^m  \, ,
\ee
where $T^m$ is a $m$-dimensional torus and $\mathcal M$ is now a transverse $9-p-d-m$ manifold. As a consequence, the metric has to be deformed as
\be
\tilde g^{(\mathcal M)}(y,q)\rmd s^2_{\mathcal M} \mapsto \sum_{\ell=1}^m \,\tilde g^{(T)}_\ell(y,q) \(\rmd \psi_\ell + i \kappa_\ell \rmd t\)^2 + g^{(\mathcal M)}(y,q)\rmd s^2_{\mathcal M} \, .
\ee
Notice that each circle of the torus has its own warp factor
\be
\tilde g^{(T)}_\ell(y,q) = R^2_\ell \, \Bigl[1+\tilde{g}^{(T)}_{2,\ell} \bigl(\,y^2+q^2\,\bigr)^2 + \ldots\Bigr] \, .
\ee
As it is clear from the formulae, one should expect this scenario to be a direct generalization of the one-dimensional case in \eqref{boostedS1metric}.

The general characteristic polynomial of the bosonic mass matrix is
\be
\mathcal Q (\lambda) = \lambda^{10-d} \(\lambda-\mu^2_T\)^d\, , \quad \mu^2_T = \frac{\beta^2_H}{4\pi^2} \(D \pt \tilde g^{(0)}_2 - \sum_{\ell=1}^m R^2_\ell \, \tilde g^{(T)}_{2,\ell}  \pt  \kappa^2_\ell \) \, ,
\ee
while the one related to the connection $A$ is the same as in \eqref{charpolArotating}. So, as expected, the above spectrum is a trivial generalization of the one described in \eqref{chargedspectrum}  for $p+d=8$ (\ie the single circle case). Then, in this case, we conclude that the map in \eqref{chargedmapping} is enough to derive $T_H$ once we make the extra substitution
\be\label{extramaptori}
R^2_{\mathcal M} \kappa^2 \mapsto \sum_{\ell=1}^m R^2_\ell \,  \kappa^2_\ell \, , \quad  R^2_{\mathcal M} \,  \tilde g^{(\mathcal M)}_2 \kappa^2 \mapsto \sum_{\ell=1}^m R^2_\ell \,  \tilde g^{(T)}_{2,\ell} \,  \kappa^2_\ell\,.
\ee
As an example, we can apply this result to a rotating string on the torus of $AdS_3 \times S^3 \times T^4$, for $p=0$, $d=2$, $m=4$.

\section{Thermal scalar derivation and specific examples}
\label{sec:effective}

In this section we study the effects of rotations, in all the types of subspaces considered in section \ref{Sec:confining_bg}, from the target space perspective, \ie using the thermal scalar effective approach. From one hand, this allows us to compute the contributions that are not accounted for by the quadratic world-sheet theory at NNLO (namely, the value of $k_2$ in the coefficient \eqref{tNNLOtNNNLO}). On the other hand, it allows us to derive general formulae for the NNNLO contributions to $T_H$.  
At LO and NLO, instead, the results of the effective approach and the sigma-model one always coincide.
Useful details for what follows can be found, \eg in \cite{Bigazzi:2024biz}, to which we address the interested readers. The thermal scalar approach in the rotating case in flat space and $AdS$ was first considered in \cite{Seitz:2025wpc,Ekhammar:2025efc}.
Here we generalize those results in various dimensions, planes of rotation and backgrounds.

A useful parameterization of the squared inverse Hagedorn temperature is
\begin{subequations}\label{nearHagnu}
    \be
\beta_H^2 = \frac{8\pi^2 \alpha'}{h} \left( 1 + c_0 \, \alpha'^{1/2} + \tilde c_1 \, \alpha'\ + \tilde c_2 \, \alpha'^{3/2} + \ldots\right) \, .
\ee
The relation among $h$, $c_0$, $\tilde c_1$ and $\tilde c_2$ with the parameters of the $\alpha'$-expansion in \eqref{expandedTH} is
\begin{align}
    &\label{c0c1}h=8\pi^2\mathfrak \pt t_{\text{LO}}^2 \, , \quad c_0 = - \frac{2 \, \mathfrak t_{\text{NLO}}}{\mathfrak t_{\text{LO}}} \, , \quad \tilde c_1=\frac{3\,\mathfrak t_{\text{NLO}}^2-2 \, \mathfrak t_{\text{LO}}\,\mathfrak t_{\text{NNLO}}}{\mathfrak t_{\text{LO}}^2} \, , \\[1ex]
    &\tilde c_2= - \frac{4\,\mathfrak t_{\text{NLO}}^3-6\,\mathfrak t_{\text{LO}}\,\mathfrak t_{\text{NLO}}\,\mathfrak t_{\text{NNLO}}+ 2 \, \mathfrak t_{\text{LO}}^2 \,\mathfrak t_{\text{NNNLO}}}{\mathfrak t_{\text{LO}}^3} \, .
\end{align}
\end{subequations}
In the following, we will take advantage of this map to fix the overall factor $h$ on a case-by-case basis for convenience. This is just a measure to declutter equations. Of course, the consistency with the results of the effective computation will be checked a posteriori. 

For the sake of simplicity, the $\log2$ terms will be often omitted in the perturbative analysis of this section. Indeed, as already stressed, the effective approach is not enough to determine them from first principles (although a general recipe to take them into account in the thermal scalar effective action has been proposed in \cite{Harmark:2024ioq} and has passed several non-trivial tests \cite{Ekhammar:2025efc}), and we would be forced to rely on arbitrary parameters all the way through. We will thus often work with a ``bare'' (i.e.~without the $\log 2$ terms) version of $\tilde c_1$ and $\tilde c_2$ which we will denote as $c_1, c_2$. 

It follows that the coefficients in the expansion of $\Delta\mathcal E$ in \eqref{gendeltaE} can be expressed in terms of the ``bare" parameters as
\be\label{effk2k3}
\boxed{k_2 = \frac{1}{2} \, g_{tt}(0) \( c_1 - \frac12 \,  c_0^2 \) \, , \quad k_3 = \frac{1}{2} \, g_{tt}(0) \, \sqrt{\frac{h}{2}} \, \[c_2 - \frac32 \, c_0 \(c_1 - \frac{5}{12} \, c_0^2\)\] \, , }
\ee
where $g_{tt}(0)$ is the ``$tt$" component of the metric in \eqref{boostedgenericmetric}, possibly deformed by the presence of twists with any direction of the internal space and evaluated at the Hagedorn point. This fills the gap in the final proposal \eqref{rotatingTH}.

Later on, the reader can restore the $\log 2$-terms by dressing the bare coefficients appropriately. The general map is
\begin{subequations}\label{genshifts}
    \begin{align}
        &\label{shiftlog}\tilde c_1 = c_1 + \frac{2}{g_{tt}(0)}  \sum_b \bar\mu^2_b \, \log2  \, , \\
&\label{shiftlog2}\tilde c_2 = c_2 + \frac{3 \, c_0}{g_{tt}(0)} \sum_b \bar\mu^2_b\, \log2 \,  ,
    \end{align}
\end{subequations}
where $\bar\mu_b$ are the bosonic world-sheet masses in units of $\beta_H/2\pi$. It relies on the interplay with the NNLO contribution from the world-sheet in \eqref{rotatingTH}, properly modified according to the prescriptions discussed in section \ref{sec:WSnoncontr} (see, \eg \eqref{chargedshifting} and \eqref{chargedmapping}). In the absence of rotations in the non-contractible cycle, the $\log 2$ terms do not (explicitly) depend on any angular velocity or any chemical potential. Therefore, at least formally,\footnote{Notice that $c_0$ may depend on some angular velocity related to rotations in the contractible cycle.} the above shifts reduce exactly to the ones reported in \cite{Bigazzi:2024biz}, namely
\begin{subequations}
\begin{align}
&\label{shiftlogbis}\tilde c_1= c_1 + 2\,d\, \tilde g_2^{(0)} \log2 \, , \\[1ex]
&\label{shiftlog2bis}\tilde c_2 = c_2 + 3 \, c_0 \,d\, \tilde g_2^{(0)}\log2 \, ,
\end{align}    
\end{subequations}
where $d$ and $\tilde g^{(0)}_2$ have been defined in \eqref{boostedgenericmetric}.

\subsection{Rotation in Minkowski}

We start our journey from the same scenario discussed in section \ref{sec:rotMinkDisk}. Here, we will focus on rotations in the Minkowskian sector of the background metric \eqref{boostedgenericmetric}. It turns out to be convenient to parameterize the $d$-dimensional contractible cycle as
\be
\rmd \vec y^2 = \rmd r^2 + r^2 \pt \tilde{g}^{(\Omega)}(r) \pt \rmd\Omega_{d-1}^2 \, ,
\ee
for some warp factor $\tilde{g}^{(\Omega)}$ depending on the radius $r^2=y^2_1 +\ldots+y^2_d$ and $\Omega_{d-1}$ being a $(d-1)$-sphere with unit radius. For notational convenience, we mostly show the results for a single plane of rotation generated by the polar coordinates $\{\rho,\theta\}$ and with a reduced angular velocity $\beta_H \pt \omega=2\pi\nu$. The general case accounting for $n_m$ planes of rotation is immediately recovered with the substitution $|\nu| \rightarrow \sum_{k=1}^{n_m} |\nu^{(m)}_k|$. Remember that the $\nu$-parameters have been introduced in equation \eqref{Dandangvel}.

In the following, we consider the effective theory to be just the quadratic action for the thermal scalar $\chi$ \cite{Horowitz:1997jc} 
\be\label{essechi}
S_{\chi} \approx \beta \, V_{\Omega_{d-1}} \, V_{\mathcal M} \hspace{-4pt} \int \hspace{-4pt} \rmd^p\vec x \, \rmd r \, \sqrt{-g} \, e^{-2\,\phi} \{ g^{pq} \, \partial_p \chi^* \, \partial_q \chi + m^2_{eff} \, \chi^* \chi\}\,,
\ee
where the effective mass is
\be
\alpha' m^2_{eff} = \frac{\beta^2  g_{tt}-8 \pi^2 \alpha'}{4\pi^2\alpha'}\, ,
\ee
with
\be 
g_{tt} = D \pt \tilde{g}^{(0)}(r) (1 + \omega^2 \rho^2)\,.
\ee 
We also assume that the dilaton does not depend on $\rho$.

In the effective approach, the Hagedorn temperature is the one at which the thermal scalar becomes massless.

With the ansatz $\chi(\vec x , r, \rho) = e^{i \vec p \cdot \vec x} \, w(r,\rho)$ and in the limit $M^2 = - \eta_{ij} p^i p^j\to0$, the equation of motion for the scalar reads
\be\label{eqw}
-\frac12 w'' + \frac12 w' \left(2\phi'-\partial_r \log{\sqrt{-g}}  \right)- \frac12 \frac{\ddot{w}}{D \tilde{g}^{(0)}} - \frac{1}{2} \frac{\dot{w}}{D \tilde{g}^{(0)}}  \left(\partial_\rho \log{\sqrt{-g}}  \right)  + \frac12 m^2_{eff} w = 0\,,
\ee
where the primes are derivatives w.r.t.~$r$ and the dots are derivatives w.r.t.~$\rho$.
We consider the expansion
\begin{subequations}
    \bea
&& \tilde{g}^{(0)}(r)= 1+\tilde{g}^{(0)}_{2} r^2 + \tilde{g}^{(0)}_{4} r^4 + ...\,, \\ 
&& \tilde{g}^{(\Omega)}(r)= 1+\tilde{g}^{(\Omega)}_{2} r^2 + ... \,, \\
&& \tilde g^{(\cal M)}(r) = R^2_{\cal M}(1+\tilde{g}^{(\cal M)}_{2} r^2 + ...)\,, \\
&& \phi (r)= \phi_0 + \phi_2 r^2+ ... \,,\\
&& \sqrt{-g} = C r^{d-1} \rho (1 + \tilde g_2 r^2 + ...)\,,
\eea
with the definitions
\begin{align}
    &C = D^{\frac{p+1}{2}} R^{9-d-p}_{\cal M}= \left(2\pi\alpha'T_s\right)^{\frac{p+1}{2}}R^{9-d-p}_{\cal M} \, ,\\
    &\tilde{g}_2 \equiv \frac12 \left((p+1)\tilde{g}^{(0)}_{2} + (d-1)\tilde{g}^{(\Omega)}_{2} + (9-d-p)\tilde{g}^{(\cal M)}_{2} \right) \,.
\end{align}
\end{subequations}

We are going to calculate the Hagedorn temperature up to NNLO in an expansion in $\alpha'$.
We do not take the small-$\nu$ limit (cf.~discussion below \eqref{genimpleqn}), \ie we do not scale $\rho$ as $r \sim \alpha'^{1/4}$.
Rather, $\rho$ keeps the natural scaling $\rho \sim \sqrt{\alpha'}$.\footnote{In section \ref{sec:rotMinkDisk}, we showed that the world-sheet modes related to the Minkowskian sector of the geometry are massless (see the bosonic mass matrix in \eqref{rotatingbosmassmat}). Therefore, the on-shell fluctuations along these directions scales as $\sqrt{\alpha'/|\nu|}$ (cf. the solutions in \eqref{genbosonicmodes}). So, keeping $\nu$ fixed in the $\alpha'$ expansion, we get the $\mathcal O(\sqrt{\alpha'})$ behavior. To the contrary, if we keep $\omega=2\pi\nu/\beta_H$ fixed in the same limit, then we have the $\mathcal O(\alpha'^{1/4})$ scaling. Let us stress that these results crucially come from the interplay with the world-sheet approach.} Moreover, it is convenient to choose
\be\label{hqui}
h=\frac{D}{1-|\nu|}
\ee
in the ansatz \eqref{nearHagnu}.
Then the equation of motion (\ref{eqw}), multiplied by $\left[1 + \tilde g_2 r^2 + \tilde g_4 r^4 + ...\right]$ to account for the expansion of the integration measure, can be expanded as
\be 
(H_0 + H_1 + H_2) w = (E_0 + E_1 + E_2) w \,,
\ee
with
\be 
E_0 = \frac{1}{\alpha'}\left(1- \frac{D \beta^2_{LO}}{8\pi^2 \alpha'} \right)\,, \qquad E_1 = -\frac{(1-|\nu|)}{\sqrt{\alpha'}}c_0\,, \qquad E_2 = - (1-|\nu|)c_1\,,
\ee 
and 
\be 
H_0 = \left[-\frac{1}{2 D} \partial_{\rho}^2 -\frac{1}{2 D \rho} \partial_{\rho} + \frac12 \rho^2 \left( \frac{D \nu^2}{\alpha'^2} \right)\right] \otimes \mathbb{1}_r\,, 
\ee 
\be 
H_1 = -\frac{1}{2 } \partial_{r}^2 -\frac{d-1}{2 r} \partial_{r} - \tilde g^{(0)}_2 r^2 \left[H_0 - \rho^2 \left( \frac{D \nu^2}{\alpha'^2} \right) \right] + \frac{1-|\nu|}{\alpha'}\tilde g^{(0)}_2 r^2 + \tilde g_2 r^2 (H_0 -E_0)\,, 
\ee
\bea 
H_2 &=& (2\phi_2 -\tilde g_2) r \partial_{r} + \tilde g^{(0)}_4 r^4 \left( \rho^2\frac{D \nu^2}{2\alpha'^2} \right) + \frac{1}{2D}\left[\partial_{\rho}^2 -\frac{1}{\rho} \partial_{\rho} \right]   \left[\tilde g^{(0)}_4 - (\tilde g^{(0)}_2)^2 \right] r^4 + \nonumber \\
&& (1-|\nu|) \left[ \frac{r^2}{\sqrt{\alpha'}}\tilde g^{(0)}_2 c_0 + \frac{r^4}{\alpha'}\tilde g^{(0)}_4 \right]  +\tilde g_2 r^2 (H_1 -E_1)+ \tilde g_4 r^4 (H_0 -E_0)\,.
\eea 

\subsubsection*{Leading Order result}
$H_0$ is the Hamiltonian of a two-dimensional harmonic oscillator with frequency
\be 
\omega_\rho = \frac{D |\nu|}{\alpha'}\,,
\ee  
and eigenfunctions
\be 
w_{n m} = w_{n\rho}\otimes w_{m r}\,, \qquad w_{n\rho}= \sqrt{2 \omega_\rho} L_{n/2}^{0}(\omega_\rho \rho^2) e^{-\frac{\omega_{\rho}}{2}\rho^2}\,, \qquad  n=0,1,...
\ee
where $L_{n/2}^{d/2-1}$ are the associated Laguerre polynomials and $w_{m r}$ is some discrete orthonormal basis.
The latter can be chosen later to be the one diagonalizing $H_1$.
The ground-state energy is
\be 
E_0^{(0)} = D E_0= \frac{D}{\alpha'}\left(1- \frac{D \beta^2_{LO}}{8\pi^2 \alpha'} \right) = \omega_\rho\,, \qquad \Rightarrow \qquad \beta^2_{LO} = \frac{8\pi^2 \alpha'}{D}(1-|\nu|)\,,
\ee 
consistently with the expansion (\ref{nearHagnu}) and (\ref{hqui}).

\subsubsection*{Next to Leading Order result}
At this order the correction to the ground-state energy is
\be 
E_1 = \langle w_0| H_1 |w_0 \rangle\,,
\ee 
where $|w_0 \rangle$ is the system ground-state.
Clearly the $H_0-E_0$ part of $H_1$ does not contribute.
Then we calculate\footnote{Note that the $\rho$-dependent part gives no contribution.}
\be 
\langle w_{0\rho}| H_1 |w_{0\rho} \rangle = -\frac{1}{2 } \, \partial_{r}^2 -\frac{d-1}{2 r} \, \partial_{r} + \frac12 \, r^2 \left[\frac{2(1-|\nu|)}{\alpha'}\tilde g^{(0)}_2 \right]\,,
\ee 
which is a $d$-dimensional harmonic oscillator with frequency
\be 
\omega_r = \sqrt{\frac{2(1-|\nu|)}{\alpha'}\tilde g^{(0)}_2}\,,
\ee
and eigenfunctions (fixing the LO base)
\be 
w_{m r}= \alpha_n L_{m/2}^{d/2-1}(\omega_r r^2) e^{-\frac{\omega_r}{2}r^2}\,, \qquad  m=0,1,..., \qquad \alpha_0^2 = \frac{2 \omega_r^{1/2}}{\Gamma[\frac{d}{2}]}\,. 
\ee
Thus the correction to the ground-state energy is
\be 
E_1 = \frac{d}{2}\omega_r\,, \qquad \Rightarrow \qquad c_0 = -\frac{d \sqrt{\tilde g^{(0)}_2}}{\sqrt{2(1-|\nu|)}}\,.
\ee 
The perturbative approach breaks down in the extremal limit $|\nu| \rightarrow 1$ at fixed $\alpha'$. 

\subsubsection*{Next to Next to Leading Order result}
At this order the correction to the ground-state energy is
\be 
E_2 = \langle w_0| H_2 |w_0 \rangle + \sum_{(n,m)\neq (0,0)} \frac{|\langle w_{n\rho} w_{m r}|H_1| w_{0\rho} w_{0 r}\rangle|^2}{E_0^{(0)} - E_0^{(n)}}\,.
\ee 
Clearly the $H_0-E_0$ parts of $H_1$ and $H_2$ do not contribute.

Let us first calculate the VEV of the part of $H_2$ without $H_1-E_1$.
Let us call it $\Delta H_2$.
It turns out that\footnote{Note that again the $\rho$-dependent part gives no contribution.}
\be 
\langle w_{0\rho}| \Delta H_2 |w_{0\rho} \rangle = (2\phi_2 -\tilde g_2) r \partial_{r} - \frac{d \sqrt{1-|\nu|}}{\sqrt{2\alpha'}}\Bigl(\tilde g^{(0)}_2 \Bigr)^{3/2} r^2 +  \left[ \frac{|\nu|}{2\alpha'} \Bigl(\tilde g^{(0)}_2\Bigr)^2 + \frac{(1-|\nu|)}{\alpha'} \tilde g^{(0)}_4 \right] r^4,
\ee 
so that calculating its VEV on $w_{0r}$ we find
\be \label{parziale}
\langle w_{0}| \Delta H_2 |w_{0} \rangle = -\frac{d}{2}(2\phi_2 -\tilde g_2) - \frac{d^2}{4} \tilde g^{(0)}_2 + \frac{d(d+2)}{8} \left[ \frac{\tilde g^{(0)}_4}{\tilde g^{(0)}_2} + \frac{\left(\tilde g^{(0)}_2\right)^2}{2 \tilde g^{(0)}_2} \frac{|\nu|}{1-|\nu|}\right]\,.
\ee
Performing the same calculation with $H_1 - E_1$ (the part of $H_2$ we were missing), we find that its contribution vanishes identically.

Then we move on to $\langle w_{n\rho} w_{m r}|H_1| w_{0\rho} w_{0 r}\rangle$.
First, note that the pure-$r$-dependent part of $H_1$ cannot contribute, since it is the identity in the $\rho$-dependent part of the Hilbert space.
Only 
\be 
- \tilde g^{(0)}_2 r^2 \left[H_0 - \rho^2 \left( \frac{D \nu^2}{\alpha'^2} \right) \right]
\ee
can contribute.
In fact, this is quadratic in both $r$ and $\rho$ (the latter just on the ground state, of course), so it can connect the ground state just with the $n=m=2$ states.
We get
\be 
\frac{|\langle w_{2 \rho} w_{2 r}|  \tilde g^{(0)}_2 r^2 \left[H_0 - \rho^2 \left( \frac{D \nu^2}{\alpha'^2} \right) \right] | w_{0\rho} w_{0 r}\rangle|^2}{E_0^{(0)} - E_0^{(2)}} = -\frac{d}{8}\frac{\left(\tilde g^{(0)}_2\right)^2}{\tilde g^{(0)}_2} \frac{|\nu|}{1-|\nu|}\,.
\ee 

Combining this with (\ref{parziale}) we get
\be 
c_1 = - \frac{E_2}{1-|\nu|} = - \frac{1}{1-|\nu|} \left[\frac{d(d+2)}{8} \frac{\tilde g^{(0)}_4}{\tilde g^{(0)}_2} + \frac{d}{2}(\tilde g_2- 2\phi_2) - \frac{d^2}{4} \tilde g^{(0)}_2 + \frac{d^2}{16} \tilde g^{(0)}_2 \frac{|\nu|}{1-|\nu|}  \right] \,.
\ee 
Consistently, in the limit $|\nu|\rightarrow 0$ the above expression for $c_1$ reduces to formula (5.33) of \cite{Bigazzi:2024biz}. 

\subsubsection{Example: Flat space}
\label{sec:effflatspace}

The first relevant example is flat space, where $D=1$, $\tilde g^{(0)}=1$, $d=0$ and there is no non-contractible manifold.
Thus $c_i=0$ and we are left with the exact known result
\be 
\beta_H^2|_{flat\, space} = 8\pi^2 \alpha'(1-|\nu|) 
	\,,
\ee   
that reproduces exactly the outcome of the world-sheet approach in \eqref{flatrotatingbetaH}. Expanding $1/\beta_H$ for small $|\nu|$ one can recover the result in \cite{Seitz:2025wpc}.\footnote{The perturbative-in-$|\nu|$ NLO results in \cite{Seitz:2025wpc} can be extended to NNLO and generalized to any background of the class we are considering by repeating the arguments of this section with the scaling of $\rho \sim r \sim \alpha'^{1/4}$.} 

\subsubsection{Example: WYM theory}

As a second example we consider the WYM background \cite{Witten:1998zw}, whose relevant parameters have been recalled in \eqref{WYMparameters}. The parameters there must be equipped with the following extra data \cite{Bigazzi:2024biz}
\be\label{wymcoeffs}
\phi_2 = \frac{27}{32 M_{KK} R^3}\,, \quad \tilde{g}^{(0)}_{4}= \frac{81}{512 (M_{KK} R^3)^2}\,, \quad \tilde{g}^{(\Omega)}_{2}= -\frac{3}{4 M_{KK} R^3}\,.
\ee
The result can be given as an expansion in the ratio of $M_{KK}$ with the string tension $T_s$,
\be 
2\pi\alpha' T_s=\frac{8 }{27}R^3 M_{KK}^3\,.
\ee
Thus we find
\be \label{beta2wym}
\beta_H^2|_{\text{WYM}} = \frac{4\pi}{T_s} (1-|\nu|)\Biggl[ 1 - \frac{M_{KK}}{\sqrt{2\pi T_s (1-|\nu|) }}+ \biggl( 2\log{2} -\frac{2-|\nu|}{8(1-|\nu|)^2} \biggl)\biggl(\frac{M_{KK}}{\sqrt{2\pi T_s}}\biggr)^{\hspace{-2pt}2} + ...\Biggr]\hspace{-1pt}.
\ee


\subsection{Rotation in the contractible cycle}

In this section we consider a single rotation in the $d$-dimensional contractible cycle of the geometry \eqref{boostedgenericmetric}, with (real) angular velocity $\Omega$.
The case of rotation in $AdS$ is included in this analysis for $p=0$.
We will focus on the cases $d=2$, $3$.
The $d=4$ case for $AdS$ (i.e.~$AdS_5$) has been analyzed in \cite{Seitz:2025wpc,Ekhammar:2025efc}.
The overall procedure is quite similar to the one of the previous section, so we skip some details.
We parameterize the inverse Hagedorn temperature as in \eqref{nearHagnu} for
\be
h=D \, .
\ee
Again, the world-sheet analysis of this scenario is contained in section \ref{sec:rotMinkDisk}.

\subsubsection{$d=2$, e.g.~global $AdS_3$}

In this case the cycle is a circle, the metric reads
\be \label{metricd2}
\rmd s^2 = D \tilde{g}^{(0)}(r) \left(\rmd t^2 + \eta_{ij} \rmd x^i \rmd x^j \right) +\rmd r^2 + r^2 \tilde{g}^{(\Omega)}(r)(\rmd \theta + i \pt \Omega \rmd t)^2 + \tilde{g}^{(\mathcal M)}(r) \rmd s_{\cal M}^2 \,,
\ee
and since the thermal scalar $w$ still depends only on $r$, there is little difference with the non-rotating case in \cite{Bigazzi:2024biz}, encoded in the dependence on $\Omega$ of 
\be 
g_{tt} = D \tilde{g}^{(0)} - r^2 \tilde{g}^{(\Omega)} \Omega^2 \,.
\ee 
The structure in (\ref{nearHagnu}) solves automatically the equation for $w$ at LO, while the equation at NLO is the one for a two-dimensional harmonic oscillator, and higher order corrections can be computed in perturbation theory. 

The $\Omega$-dependent part of $g_{tt}$ enters as an extra-term in $m_{eff}$ from NLO, where it gives the contribution
\be 
-\frac{\beta^2 \Omega^2 \tilde{g}^{(\Omega)} r^2}{8\pi^2 \alpha'^2} \,.
\ee1+3
This just shifts the frequency of the two-dimensional harmonic oscillator to
\be 
\omega_0 = \sqrt{\frac{2}{\alpha'}} \sqrt{\tilde{g}^{(0)}_2 -\frac{\Omega^2}{D}}\,,
\ee  
so that
\be 
c_0 = - \sqrt{2} \sqrt{\tilde{g}^{(0)}_2 -\frac{\Omega^2}{D}}\,.
\ee 
Notice that the limiting case $\Omega^2 \rightarrow D \tilde{g}^{(0)}_2$ is singular, since the frequency of the harmonic oscillator vanishes, the solution ``de-localizes'' \cite{Ekhammar:2025efc}.
This phenomenon is present for any $d$. The limiting case can be treated separately as in \cite{Ekhammar:2025efc}.

At NNLO the rotation provides an extra-term to the perturbation Hamiltonian which reads 
\be 
-\frac{\Omega^2 r^2}{D \alpha'} \left(c_0 \sqrt{\alpha'}+r^2 \tilde{g}^{(\Omega)}_2 \right) \subset H_1\,.
\ee
Evaluating 
\be 
\langle w_0 |H_1 |w_0 \rangle\,,
\ee 
with the extra term and the above expressions of $\omega_0$ and $c_0$, gives the result
\be  
c_1 = (2\phi_2-\tilde g_2) + \tilde{g}^{(0)}_2 - \frac{\tilde{g}^{(0)}_4}{\tilde{g}^{(0)}_2 -\Omega^2/D} - \frac{\Omega^2}{D}\left(1- \frac{\tilde{g}^{(\Omega)}_2}{\tilde{g}^{(0)}_2 -\Omega^2/D} \right)\,.
\ee 

\subsubsection{Example: WYM theory}

Let us consider again the WYM example with parameters in \eqref{WYMparameters} and \eqref{wymcoeffs}, the cycle where the rotation takes place being the circle of the cigar.
The result reads 
\be 
\beta_H^2|_{\text{WYM}} = \frac{4\pi}{T_s}\Biggl[ 1 - \frac{M_{KK}}{\sqrt{2\pi T_s  }}\sqrt{1-\frac{2\Omega^2}{M_{kk}^2}}+ \Biggl[ 2\log{2} -\frac{1}{4}\Biggl( \frac{4\Omega^2}{M_{kk}^2}+ \frac{1}{1-\frac{2\Omega^2}{M_{kk}^2}}  \Biggr) \Biggr] \hspace{-2pt}\Biggl(\frac{M_{KK}}{\sqrt{2\pi T_s}}\Biggr)^{\hspace{-4pt}2} \hspace{-2pt} + \ldots\Biggr].
\ee

\subsubsection{NNNLO for global $AdS_3$}

Here and in the following subsections we will present the NNNLO results only for the global-$AdS$ cases. The only, pragmatic, reason is that, for general non-$AdS$ confining backgrounds, it is unlikely that subleading corrections to the Hagedorn temperature at strong coupling, can be computed in the near future using complementary field theory methods in order to cross-check the holographic predictions.\footnote{Notice that the leading order string corrections to the type IIA and type IIB supergravity actions are of $\mathcal O(\alpha'^3)$. Hence these corrections can be safely neglected in general if we stick with the Hagedorn temperature expansion to NNNLO. This remark revises, to some extent, certain claims made in previous studies.} The NNNLO corrections to the Hagedorn temperature of general holographic confining theories at zero chemical potentials have been computed in \cite{Bigazzi:2024biz}.

Borrowing data from \cite{Bigazzi:2024biz}, in the $AdS_{3}$ case we have $D=1$ and, setting to one the $AdS$ radius,
\begin{subequations}\label{AdS3coeff}
    \begin{align}
         &\tilde{g}^{(0)}_2 =1\,, \quad  \tilde{g}^{(0)}_4 =\frac13\,, \quad  \tilde{g}^{(0)}_6 =\frac{2}{45}\,, \quad  \tilde{g}_2 =\frac23\,,\\
         &\tilde{g}_4 =\frac{2}{15}\,, \quad \tilde{g}^{(\Omega)}_2 =\frac13\,, \quad \tilde{g}^{(\Omega)}_4 =\frac{2}{45}\,, \quad \phi_i=0\,.
    \end{align}
\end{subequations}
Thus in this case we have
\be 
c_0 = -\sqrt{2} \sqrt{1-\Omega^2}\,, \quad c_1= -\Omega^2\,, \quad \omega_0 =\sqrt{\frac{2}{\alpha'}} \sqrt{1 -\Omega^2}\,.
\ee
The relevant operators to consider at NNNLO are
\be 
\Delta H_1 = -\frac23 r \partial_r + \frac{c_0}{\sqrt{\alpha'}}r^2 + \frac{1}{3\alpha'}r^4 - \frac{\Omega^2 c_0}{\sqrt{\alpha'}}r^2 - \frac{\Omega^2}{3\alpha'}r^4 \,, 
\ee
with $E_1 = \Omega^2$, and 
\be 
\Delta H_2 = \frac{8}{45} r^3 \partial_r + c_1 r^2 + \frac{c_0}{3\sqrt{\alpha'}}r^4 + \frac{2}{45\alpha'}r^6 - \Omega^2 c_1 r^2- \frac{\Omega^2 c_0}{3\sqrt{\alpha'}}r^4 - \frac{2\Omega^2}{45\alpha'}r^6 \,. 
\ee
Since $\Delta H_1$ has only quadratic and quartic terms (possibly after partial integration), only the eigenfunctions with $n=2, 4$ ($w_{2,4}$, Laguerre functions in two dimensions) contribute, so we have to evaluate
\be 
-\sqrt{\alpha'}c_2 = E_2 = \langle w_0| \Delta H_2 + \frac23 r^2(\Delta H_1 -E_1)|w_0 \rangle +  \frac{|\langle w_{2} |\Delta H_1| w_{0}\rangle|^2}{-2 \omega_0} + \frac{|\langle w_{4} |\Delta H_1| w_{0}\rangle|^2}{-4 \omega_0} \,.
\ee 
This gives
\be 
c_2|_{AdS_3} = \frac{3-2\Omega^4}{4 \sqrt{2}\sqrt{1-\Omega^2}}\,,
\ee 
which has the correct limit for $\Omega \rightarrow 0$ \cite{Bigazzi:2024biz}.
The complete result for the spin dependence of the Hagedorn temperature, including the shifts (\ref{shiftlog}), (\ref{shiftlog2}) reads
\begin{align}
    \beta_H^2|_{AdS_3} = 8\pi^2 \alpha'\Biggl[ 1 &-\sqrt{2} \sqrt{1-\Omega^2}  \, \alpha'^{1/2} +\left(4 \log{2}-\Omega^2 \right) \, \alpha'\ + \\
    &+\left( \frac{3-2\Omega^4}{4 \sqrt{2}\sqrt{1-\Omega^2}} - 6 \sqrt{2}\sqrt{1-\Omega^2} \log{2}\right) \alpha'^{3/2} + ...\Biggr] 
\,.\nonumber
\end{align}
\subsubsection{$d=3$, e.g.~global $AdS_4$}

In this case the metric reads
\begin{align}\label{metricd3}
\rmd s^2 = D \tilde{g}^{(0)}(r) &\left(\rmd t^2 + \eta_{ij} \rmd x^i \rmd x^j \right) + \tilde{g}^{(\mathcal M)}(r) \rmd s_{\cal M}^2 + \\
&+\rmd r^2 + r^2 \tilde{g}^{(\Omega)}(r)\left[\rmd \theta^2 + \sin^2{\theta}(\rmd \varphi + i\,\Omega \rmd t)^2 \right]  \,,\nonumber
\end{align}
so
\be 
g_{tt} = D \tilde{g}^{(0)} - r^2 \tilde{g}^{(\Omega)}\sin^2{\theta} \,\Omega^2 \,.
\ee 
Now the thermal scalar depends on two variables, $w=w(r,\theta)$, which get entangled at NLO, so it is convenient to change them to \cite{Seitz:2025wpc,Ekhammar:2025efc}
\be 
\rho = r \sin{\theta}\,, \qquad z = r \cos{\theta}\,,
\ee
giving
\begin{align}\label{metricd3b}
\rmd s^2 = D \tilde{g}^{(0)}(r) &\left(\rmd t^2 + \eta_{ij} \rmd x^i \rmd x^j \right) +\frac{(\rho \rmd\rho + z \rmd z)^2}{\rho^2 + z^2} + \\
&+\tilde{g}^{(\Omega)} \frac{(z \rmd\rho - \rho \rmd z)^2}{\rho^2 + z^2} + \tilde{g}^{(\Omega)}\rho^2(\rmd \varphi + i\pt\Omega \rmd t)^2 + \tilde{g}^{(\mathcal M)}(r) \rmd s_{\cal M}^2 \,,\nonumber
\end{align}
and 
\be 
g_{tt} = D \tilde{g}^{(0)} -  \tilde{g}^{(\Omega)}\rho^2 \Omega^2\,. 
\ee
The kinetic terms in the equation for $w$ are
\begin{subequations}
    \begin{align}
        & -\left[\partial_\rho \log{\left(\sqrt{-g}e^{-2\phi}\right)} \right]\left[g^{\rho\rho}\partial_\rho + g^{\rho z}\partial_z \right] - \left[\partial_z \log{\left(\sqrt{-g}e^{-2\phi}\right)} \right]\left[g^{zz}\partial_z + g^{\rho z}\partial_\rho \right] \\
        & - \left[ g^{\rho\rho}\partial^2_\rho + g^{zz}\partial^2_z + \left(g^{\rho\rho}\partial_\rho + g^{\rho z}\partial_z \right) \partial_\rho + \left( g^{zz}\partial_z + g^{\rho z}\partial_\rho\right) \partial_z +2 g^{\rho z}\partial_\rho \partial_z \right]\,,
    \end{align}
\end{subequations}
while the effective mass is
\bea
&& \frac12 \,m^2_{eff}  =  -\frac{2}{\alpha'} + \frac{1}{D \alpha'} \left[ 1+ c_0 \sqrt{\alpha'} + c_1 \alpha' + c_2 \alpha'^{3/2}+\ldots \right] \times \\
&& \qquad \qquad\qquad\qquad\times \Biggl[D\left( 1 + \tilde g_2^{(0)} (\rho^2+z^2) + \tilde g_4^{(0)} (\rho^2+z^2)^2 + \ldots\right) + \nonumber\\
&&\quad\qquad\qquad\qquad\qquad- \rho^2 \Omega^2 \left( 1 + \tilde g_2^{(\Omega)} (\rho^2+z^2) + \tilde g_4^{(\Omega)} (\rho^2+z^2)^2 + \ldots\right) \Biggr]\,. \nonumber
\eea

At NLO the full operator reads
\be 
H_0= -\frac12 \partial^2_\rho -\frac{1}{2\rho} \partial_\rho -\frac12 \partial^2_w + \frac12 \omega^2_\rho\rho^2 + \frac12 \omega^2_z z^2 + \frac{c_0}{\sqrt{\alpha'}}\,,
\ee
with
\be 
\omega_\rho = \sqrt{\frac{2}{\alpha'}} \sqrt{\tilde{g}^{(0)}_2 -\frac{\Omega^2}{D}}\,, \qquad \omega_z = \sqrt{\frac{2}{\alpha'}} \sqrt{\tilde{g}^{(0)}_2}\,.
\ee 
We thus have a two-dimensional oscillator with frequency $\omega_\rho$ and a one-dimensional oscillator with frequency $\omega_z$, so the ground-state energy is $E=-c_0/\sqrt{\alpha'}= \omega_\rho + \omega_z/2$, \ie
\be  
c_0 = - \sqrt{2} \left( \sqrt{\tilde{g}^{(0)}_2 -\frac{\Omega^2}{D}} + \frac12 \sqrt{\tilde{g}^{(0)}_2} \right)\,.
\ee

At NNLO the operator is
\begin{align}\label{H14}
H_1 = & \frac12 \,  \tilde g_2^{(\Omega)} \biggl[ z^2 \partial^2_\rho + \rho^2 \partial^2_z  - 2 z \rho \partial_\rho \partial_z - \rho \partial_\rho - z \partial_z +\\[-1ex]
&\qquad\qquad\qquad\qquad\qquad\quad- 2 \left(\tilde g_2 - 2 \phi_2 \right)\left(\rho \partial_\rho + z \partial_z \right) + \frac{1}{\rho} \left( z^2 \partial_\rho - \rho z \partial_z\right)\biggr]+\nonumber\\[1ex]
&+ \frac{c_0}{\sqrt{\alpha'}} \biggl[ \tilde g_2^{(0)} (\rho^2+z^2) - \frac{\Omega^2}{D}\rho^2 \biggr] + \frac{1}{\alpha'} \biggl[ \tilde g_4^{(0)} (\rho^2+z^2)^2 - g_2^{(\Omega)}\frac{\Omega^2}{D}\rho^2 (\rho^2+z^2) \biggr], \nonumber
\end{align}
with $E_1=-c_1$. Evaluating
\be 
E_1 = \langle w_0 | H_1 |w_0 \rangle \,,
\ee 
gives the result
\begin{eqnarray}
c_1 &=& - \frac{1}{8 D^2 \tilde g_2^{(0)} \left( \tilde g_2^{(0)} - \frac{\Omega^2}{D} \right) }\{ -8 \tilde g_2^{(0)} \Omega^4 + D^2 \left[12 (\tilde g_2-2\phi_2) \left(\tilde g_2^{(0)}\right)^2 + 4 g_2^{(\Omega)} \left(\tilde g_2^{(0)}\right)^2 +\right. \right. \nonumber \\
&& \left. \left. - 10 \left(\tilde g_2^{(0)}\right)^3 + 11 \tilde g_2^{(0)} \tilde g_4^{(0)} -4 \sqrt{\tilde g_2^{(0)}} \sqrt{\tilde g_2^{(0)} - \frac{\Omega^2}{D}} \left(\tilde g_2^{(\Omega)} \tilde g_2^{(0)} +2 \left(\tilde g_2^{(0)}\right)^2 - \tilde g_4^{(0)}\right)\right] +\right. \nonumber \\
&& \left. - D \Omega^2 \left[ 3 \tilde g_4^{(0)} - 2 \tilde g_2^{(0)} \left( 9 \tilde g_2^{(0)} - 6 (\tilde g_2-2\phi_2)  - 6 \tilde g_2^{(\Omega)} + 4 \sqrt{\tilde g_2^{(0)}} \sqrt{\tilde g_2^{(0)} - \frac{\Omega^2}{D}} \right) \right] \}   \,. \nonumber
\end{eqnarray} 

\subsubsection{NNNLO for global $AdS_4$}

Here, we present the NNNLO result for the $AdS_4$ case (for example the ABJM theory \cite{Aharony:2008ug}).
Setting to one the $AdS$ radius, the ingredients we need in this case are \cite{Bigazzi:2024biz}
\begin{subequations}\label{AdS4coeff}
    \begin{align}
        &D=1 \, , \quad \tilde{g}^{(0)}_2 =1\,, \quad  \tilde{g}^{(0)}_4 =\frac13\,, \quad  \tilde{g}^{(0)}_6 =\frac{2}{45}\,, \quad  \tilde{g}_2 =\frac56\,,\\
        &\tilde{g}_4 =\frac{91}{360}\,, \quad \tilde{g}^{(\Omega)}_2 =\frac13\,, \quad \tilde{g}^{(\Omega)}_4 =\frac{2}{45}\,, \quad \phi_i=b_i=0\,,
    \end{align}
\end{subequations}
and, using the results of the previous section,
\begin{subequations}
    \begin{align}
        &c_0 = - \sqrt{2} \left( \sqrt{1-\Omega^2} + \frac12  \right)\,, \quad c_1= -\frac58 - \Omega^2 + \sqrt{1-\Omega^2}\,, \\
        &\omega_\rho =\sqrt{\frac{2}{\alpha'}} \sqrt{1 -\Omega^2}\,,\quad \omega_z =\sqrt{\frac{2}{\alpha'}}\,.
    \end{align}
\end{subequations}
The relevant operators are $\Delta H_1 = H_1$ in (\ref{H14}) 
and
\begin{align}
    \Delta H_2 = -\frac{1}{30}(\rho^2&+z^2) \left[ z^2 \partial^2_\rho + \rho^2 \partial^2_z  - 2 z \rho \partial_\rho \partial_z - \frac{20}{3}\rho \partial_\rho -\frac{20}{3} z \partial_z + \frac{1}{\rho} \left( z^2 \partial_\rho - \rho z \partial_z\right)\right] + \nonumber \\
& + c_1 \left[ \rho^2+z^2 - \Omega^2 \rho^2 \right] +  \frac{c_0}{3\sqrt{\alpha'}}  \left[ (\rho^2+z^2)^2 - \Omega^2\rho^2 (\rho^2+z^2) \right] +\nonumber\\
&+\frac{2}{45\alpha'}\left[ (\rho^2+z^2)^3 - \Omega^2\rho^2 (\rho^2+z^2)^2 \right]\,. 
\end{align}
Since $\Delta H_1$ has only terms involving $\rho$ and $z$ at the quadratic and (total) quartic level, only the eigenfunctions with $n=0, 2, 4$ ($w_{20}, w_{02}, w_{22}, w_{40}, w_{04}$, Laguerre functions in two dimensions) contribute, so we have to evaluate
\begin{align}
     -\sqrt{\alpha'}c_2 = E_2 = \langle w_0| \Delta H_2 + \frac56 (\rho^2+z^2&)(\Delta H_1 -E_1)|w_0 \rangle +  \frac{|\langle w_{22} |\Delta H_1| w_{00}\rangle|^2}{-2 (\omega_\rho + \omega_z)}+ \\
 & +\frac{|\langle w_{20} |\Delta H_1| w_{00}\rangle|^2}{-2 \omega_\rho } + \frac{|\langle w_{02} |\Delta H_1| w_{00}\rangle|^2}{-2  \omega_z}+\nonumber\\
 &+  \frac{|\langle w_{40} |\Delta H_1| w_{00}\rangle|^2}{-4 \omega_\rho} +  \frac{|\langle w_{04} |\Delta H_1| w_{00}\rangle|^2}{-4 \omega_z}\,.\nonumber
\end{align} 
This gives
\be 
c_2|_{AdS_4} = \frac{47+55\sqrt{1-\Omega^2}+\Omega^2 \left(1+ 8\sqrt{1-\Omega^2} -8\Omega^2(5+2\sqrt{1-\Omega^2}) \right)}{32 \sqrt{2}\left(1-\Omega^2+\sqrt{1-\Omega^2}\right)}\,,
\ee 
which has the correct limit for $\Omega\rightarrow 0$ \cite{Bigazzi:2024biz}.
The complete result for the spin dependence of the Hagedorn temperature reads
\begin{align}
    \hskip -10pt\beta_H^2|_{AdS_4} = 8\pi^2 \alpha'\Biggl[ & 1 - \sqrt{2} \left( \sqrt{1-\Omega^2} + \frac12  \right) \, \alpha'^{\frac12} +\left(6\log{2} -\frac58 - \Omega^2 + \sqrt{1-\Omega^2} \right)  \, \alpha' +\nonumber\\ 
	& + \Biggl( \frac{47+55\sqrt{1-\Omega^2}+\Omega^2 \left(1+ 8\sqrt{1-\Omega^2} -8\Omega^2(5+2\sqrt{1-\Omega^2}) \right)}{32 \sqrt{2}\left(1-\Omega^2+\sqrt{1-\Omega^2}\right)} + \nonumber \\
    &\qquad\qquad\qquad\qquad\qquad- 9\sqrt{2}\left(\sqrt{1-\Omega^2}+\frac{1}{2} \right)\log{2}\Biggr) \alpha'^{3/2} + \ldots\Biggr] 
\,.
\end{align}

\subsubsection{Notes on the general case}

For generic $d$ and generic background the expressions can be deduced as well, but the formulas tend to become quite cumbersome.
At the NLO the general formula for multiple rotating directions is instead quite simple
\be  
c_0 = - \sqrt{2} \left(\sum_{i=1}^{[\frac{d}{2}]} \sqrt{\tilde{g}^{(0)}_2 -\frac{\Omega^2_i}{D}} + \frac12 \sqrt{\tilde{g}^{(0)}_2}\delta_{d,odd} \right)\,.
\ee
It corresponds exactly to $c_0=-2\,\mathfrak t_{\text{NLO}}/\mathfrak t_{\text{LO}}$ with $\mathfrak t_{\text{LO}}$ and $\mathfrak t_{\text{NLO}}$ given in \eqref{rotatingTH} (for $n_m=0$). In $AdS_{d+1}$, where $\tilde{g}^{(0)}_2=D=1$, one gets
\be  
c_0 = - \sqrt{2} \left(\sum_{i=1}^{[\frac{d}{2}]} \sqrt{1-\Omega_i^2} + \frac12 \delta_{d,odd} \right)\,,
\ee
consistently with all the known results.

Moreover, for $AdS_{d+1}$ backgrounds the formula at the NNLO is quite compact. Considering the previous results and the one for $AdS_5$ in \cite{Ekhammar:2025efc}, the general formula is extrapolated to be
\be 
\tilde c_1 = -\frac{(8-d)(d-2)}{8} -\sum_{i=1}^{[\frac{d}{2}]} \Omega_i^2 + (d-2) \Pi_{i=1}^{[\frac{d}{2}]}\sqrt{1-\Omega_i^2} + 2\,d\,  \log2\,.
\ee


\subsection{Rotation in the non-contractible $q$-cycle}
In this section we consider a string rotating in the non-contractible cycle $\mathcal M$ of the geometry in \eqref{boostedgenericmetric}, assuming it has dimension $q$ for simplicity. In particular, we will consider a single twist with one of the directions of the internal space. 
In the following, $\kappa$ will denote the chemical potential that couples to some global $U(1)$ charge in the dual QFT. From a (reduced) bulk perspective, this corresponds to a background gauge field
\be
A_\mu \rmd z^\mu = i \pt \kappa \pt \rmd t \, .
\ee
Now, the convenient parameterization of the inverse Hagedorn temperature is \eqref{nearHagnu} equipped with
\be\label{hNonContr}
h=D - R^2_{\cal M} \kappa^2\,.
\ee
The shifts of $c_1, c_2$ due to the sigma-model $\log 2$ terms will in general be given by eqns \eqref{genshifts}.


Below, we first consider the simpler $q=1$ setup and then move to the general case.

\subsubsection{$q=1$}

This is the case of a circle, discussed in \eqref{boostedS1metric} from a world-sheet perspective.
The metric is
\be \label{metricq1}
\rmd s^2 = D \tilde{g}^{(0)}(r) \left(\rmd t^2 + \eta_{ij} \rmd x^i \rmd x^j \right) +\rmd r^2 + r^2 \tilde{g}^{(\Omega)}(r)\rmd \Omega_{d-1} + \tilde{g}^{(\mathcal M)}(r)   (\rmd \theta + i \pt \kappa \pt \rmd t)^2   \,,
\ee
and since the scalar $w$ still depends only on $r$, there is little difference with the non-rotating case in \cite{Bigazzi:2024biz}, encoded in the expression of 
\be 
g_{tt} = D \tilde{g}^{(0)} - \tilde{g}^{({\cal M})} \kappa^2 \,.
\ee 
The $\kappa$-dependent terms enter already at LO, resulting in the shift
\be 
D \rightarrow D\left(1- \frac{R^2_{\cal M} \kappa^2}{D} \right)\,.
\ee
The form of the parameterization \eqref{hNonContr} reflects this fact. Moreover, the extra term is going to enter the calculation as a shift
\be 
\tilde g^{(0)}_n \rightarrow \tilde g^{(0)}_n - \frac{R^2_{\cal M}\kappa^2}{D}\tilde{g}^{({\cal M})}_n \equiv N_n\,.
\ee
It can be explicitly verified that the two shifts combine in such a way that the $c_i$ are exactly the ones calculated in \cite{Bigazzi:2024biz} but for the replacements\footnote{Note that there is a typo in the last line of formula (5.38) in \cite{Bigazzi:2024biz}, where it should be $\phi_4 \rightarrow 2\phi_4$.} 
\be 
\tilde g^{(0)}_n \rightarrow  \frac{N_n}{1- \frac{R^2_{\cal M}\kappa^2}{D}}\,.
\ee
This is the same conclusion we reached in \eqref{chargedmapping}. It is clear that the extension to $m$ different circles (e.g.~$m$-tori $T^m$), possibly with different warp factors, is given in a straightforward way by \eqref{extramaptori} generalizing the subscript $2$ to generic values of $n$.

\subsubsection{Example: rotation in the torus of global $AdS_3 \times S^3 \times T^4$}

As an example, let us consider global $AdS_3 \times S^3 \times T^4$ (with zero $B$-filed, unit radii, and with $\tilde g^{(T)}_n=0$) and rotate on the torus.
Then, making use of the shifts defined above and the results in \cite{Bigazzi:2024biz} for $p=0$, $d=2$, with the other coefficients introduced in \eqref{AdS3coeff}, we have
\be 
\beta_H^2 = \frac{8\pi^2 \alpha'}{\left(1 - \sum_{\ell=1}^4 \kappa^2_\ell \right)} \left( 1 + c_0 \, \alpha'^{1/2} + \tilde c_1 \, \alpha'\ + \tilde c_2 \, \alpha'^{3/2} + ...\right) 
\,,
\ee
with (including the $\log 2$ terms from the sigma model)
\bea 
&& c_0 =  -\frac{\sqrt{2}}{\sqrt{1-\sum_{\ell=1}^4 \kappa^2_\ell }}\,, \qquad \tilde c_1 = \frac{4 \log{2} + \sum_{\ell=1}^4 \kappa^2_\ell }{1-\sum_{\ell=1}^4 \kappa^2_\ell }\,, \\[2ex]
&& \tilde c_2 = \frac{\sqrt{1-\sum_{\ell=1}^4 \kappa^2_\ell}}{4\sqrt{2}}  +  \frac{1}{\sqrt{2\(1-\sum_{\ell=1}^4 \kappa^2_\ell\)}} - \frac{1}{\[2\left(1-\sum_{\ell=1}^4 \kappa^2_\ell \right)\]^{3/2}} (1+24\log2)\,.\nonumber
\eea

\subsubsection{Example: rotation along the Sasaki-Einstein Reeb vector}

The $q=1$ results, with constant $\tilde{g}^{(\mathcal M)}$, can be used to fill the gap of section \ref{sec:reeb} about the dependence of the Hagedorn temperature on the chemical potential for the $U(1)_R$ R-symmetry of CFTs on spheres.\footnote{There, we focused on global $AdS_5 \times X^5$ cases, with $X^5$ being a five-dimensional SE manifold. Nevertheless, the reasoning can be extended to more general $AdS\times $SE cases.} Indeed, let us stress that the $k_{2,3}$-terms in \eqref{reebTH} cannot be fixed by the sigma-model approach. Rather, they can be computed as in \eqref{effk2k3}.  

\subsubsection{$q>1$}
We assume that the metric can be written as (\eg if the cycle is a sphere)
\begin{align}\label{metricq2}
\rmd s^2 = D \pt \tilde{g}^{(0)}(r) &\left(\rmd t^2 + \eta_{ij} \rmd x^i \rmd x^j \right) +\rmd r^2 + r^2 \pt \tilde{g}^{(\Omega)}(r) \, \rmd \Omega_{d-1}^2 +\\
&+\tilde{g}^{(\mathcal M)}(r)   \left(\rmd\theta^2 +\sin^2{\theta} \, (\rmd\varphi + i \pt \kappa \rmd t)^2 +\cos^2{\theta}  \rmd\Omega_{q-2}^2 \right)  \,. \nonumber
\end{align}
From a world-sheet perspective, the reader can find a complementary discussion about this scenario in section \ref{sec:spinonSn}. We perform the change of variable $z=\cos{\theta}$, giving
\begin{align}\label{metricq2b}
\rmd s^2 = D \tilde{g}^{(0)}(r) &\left(\rmd t^2 + \eta_{ij} \rmd x^i \rmd x^j \right) + \rmd r^2 + r^2 \tilde{g}^{(\Omega)}(r)\rmd \Omega_{d-1}^2 + \\
&+\tilde{g}^{(\mathcal M)}(r)   \left(\frac{dz^2}{1-z^2} +(1-z^2) (\rmd\varphi + i \pt \kappa \rmd t)^2 +z^2 \rmd\Omega_{q-2}^2\right)   \,,    
\end{align}
so the determinant of the metric has a power of $z^{q-2}$ and
\be 
g_{tt} = D \tilde{g}^{(0)} - (1-z^2)\tilde{g}^{(M)} \kappa^2 \,.
\ee 
We will concentrate on the strict $|\kappa| > 0$ regime for simplicity.
In this situation the dependence of the thermal scalar on $z$ has the consequence that on top of the $d$-dimensional oscillator (in $r$) as before, there is also a $(q-1)$-dimensional oscillator in $z$ with frequency\footnote{In the $\kappa \rightarrow 0$ limit the system cannot be approximated as a harmonic oscillator in $z$ (its mass term would be vanishingly small, so the wave-function would not be exponentially peaked in the (compact) direction $z$), and the required analysis is more involved  \cite{Ekhammar:2025efc}. So, we are not going to recover the known results at $\kappa=0$ from the formulas of this section.
Note that, instead, all the sigma-model contributions, from LO to NNLO, have a consistent $\kappa \rightarrow 0$ limit.}
\be 
\omega_z^2 = - \frac{2}{\alpha'} \frac{\nu^2}{D\left(1-\frac{R^2_{\cal M}\kappa^2}{D} \right)}\,.
\ee
Similarly, in \eqref{chargedspectrum} we had $(q-1)$ extra massive modes in the world-sheet spectrum.
This gives the same LO result as before, and a NLO term determined by the energy of the two oscillators, resulting in 
\be
c_0 = -\frac{d}{\sqrt{2}}\frac{\sqrt{N_2}}{\sqrt{1-R^2_{\cal M}\kappa^2/D}} -  \frac{(q-1) |\kappa|  }{\sqrt{2D}\sqrt{1-R^2_{\cal M}\kappa^2/D}}\,.
\ee
At NNLO and NNNLO we have to consider also the kinetic term correction in $z$ coming from $g^{zz}$, apart from the usual ones in $r$ and the potential from $g_{tt}$.
The results can be written explicitly but are very lengthy and not illuminating.\footnote{We explicitly reproduced the result in \cite{Ekhammar:2025efc} for $AdS_5 \times S^5$ ($d=4$, $q=5$).}

\subsubsection{Example: rotation in the sphere of global $AdS_3 \times S^3 \times T^4$}

As an explicit example, let us consider again the global $AdS_3 \times S^3 \times T^4$ background, but let the rotation be on the three-sphere (so $p=0$, $d=2$, $q=3$, constant $\tilde g^{(\mathcal M)}=1$ and other coefficients introduced in \eqref{AdS3coeff}).
One gets 
\be 
\beta_H^2 = \frac{8\pi^2 \alpha'}{1- \kappa^2} \left( 1 + c_0 \, \alpha'^{\frac12} + \tilde c_1 \, \alpha'\ + \tilde c_2 \, \alpha'^{3/2} + ...\right) 
\,,
\ee
with
\begin{align}
    &c_0 = - \sqrt{2} \sqrt{\frac{1+|\kappa|}{1-|\kappa|}}\,, \qquad \tilde c_1 = \frac{1+|\kappa| }{1-|\kappa|}+ \frac{4 (1+\kappa^2)\log{2}}{1-\kappa^2}\,, \\
    &\tilde c_2 = \frac{1+ |\kappa|\left(|\kappa| -4 \right)}{4 \sqrt{2} \, |\kappa| \left(\frac{1-|\kappa|}{1+|\kappa|} \right)^{3/2}} -6\sqrt{2}
\frac{1+\kappa^2}{\sqrt{1+|\kappa|}\left(1-|\kappa| \right)^{3/2}}
\log{2}\,.\nonumber
\end{align} 
 
\subsubsection{Example: WYM theory}

As another example, let us consider the WYM theory, the cycle where the rotation takes place being the four-sphere ($p=3$, $d=2$, $q=4$).
The radius is $R^2_{\cal M}=R^2_{S^4}= R^{3/2}\sqrt{u_0}=2 M_{KK}R^3/3$. The other parameters can be found in \eqref{WYMparameters} and \eqref{wymcoeffs}.
The result up to NNLO reads
\begin{align}
   &\beta_H^2|_{\text{WYM}} = \frac{4\pi}{T_s \left(1-\frac{9\kappa^2}{4M_{KK}^2} \right) } \{ 1 - \frac{1}{\sqrt{1-\frac{9\kappa^2}{4M_{KK}^2}}} \left(\sqrt{1-\frac{3\kappa^2}{4M_{KK}^2}} + \frac{3\kappa}{\sqrt{2}M_{KK}}\right)  \left(\frac{M_{KK}}{\sqrt{2\pi T_s  }}\right)+ 
\right. \nonumber \\ 
& \left. + \left[ \frac{2\left(1+\frac{9\kappa^2}{4M_{KK}^2} \right)\log{2}}{\left(1-\frac{9\kappa^2}{4M_{KK}^2} \right)} 
+\frac{\frac{7}{12}+\frac{\kappa \left(27\kappa-14\sqrt{8M_{KK}^2-6\kappa^2} \right)}{24 M_{KK}^2}-\frac{3\kappa^3 \left(21\kappa-4\sqrt{8M_{KK}^2-6\kappa^2} \right)}{64 M_{KK}^4}}{\left(1-\frac{9\kappa^2}{4M_{KK}^2} \right)\left(1-\frac{3\kappa^2}{4M_{KK}^2} \right)}
\right]\hspace{-4pt}\left(\frac{M_{KK}}{\sqrt{2\pi T_s}}\right)^{\hspace{-2pt}2} 
\}.
\end{align}

\subsubsection{Example: ABJM}

The ABJM case is slightly different since the metric of the internal manifold ($\CP^3$) is not exactly in the form (\ref{metricq2}). The world-sheet approach to this scenario is included in section \ref{sec:WSCpn}, specifying those results for $n=3$.
A commonly used parameterization (including our shift) is\footnote{Notice that here we double the background gauge field to match the notations used in section \ref{sec:WSCpn}.}
\begin{align}
    \rmd s^2_{\CP^3} = &\rmd\mu^2 + \frac14 \cos^2{\mu}\, \rmd\Omega_{2,1}^2+ \frac14 \sin^2{\mu}\, \rmd\Omega_{2,2}^2 + \\ 
    & + \sin^2{\mu}\cos^2{\mu}\,\biggl(\rmd\psi + 2 \pt i \pt \kappa \rmd t + \frac12\cos{\theta_1} d\phi_1 -\frac12 \cos{\theta_2} \rmd \phi_2  \biggr)^{\hspace{-2pt}2} \,,\nonumber
\end{align}
where $\rmd\Omega_{2,1}^2$ ($\rmd\Omega_{2,2}^2$) is the round metric of a unit $S^2$ described by the polar angle $\theta_1$ ($\theta_2$) and the azimuthal angle $\phi_1$ ($\phi_2$).
We are going to work in units where $\ell_{AdS}=1$, so that the ten-dimensional metric is \cite{Aharony:2008ug}
\be
\rmd s^2_{10} =\rmd s^2_{AdS_4} + R^2_{\mathcal M} \rmd s^2_{\CP^3} \, , \quad R^2_{\mathcal M} = 4 \,.
\ee
So, defining $z=\cos{2\mu}$ we have
\begin{align}
    4\,\rmd s^2_{\CP^3} = &\frac{\rmd z^2}{1-z^2} + \frac12 (1+z)\, \rmd\Omega_{2,1}^2+ \frac12 (1-z)\, \rmd\Omega_{2,2}^2 + \\
    &+(1-z^2)\left(\rmd\psi + 2 \pt i \pt \kappa \rmd t + \frac12\cos{\theta_1} \rmd\phi_1 -\frac12 \cos{\theta_2} \rmd\phi_2  \right)^{\hspace{-2pt}2}\,.\nonumber
\end{align}
Then we can proceed as before, obtaining at LO the correct form of the parameterization in \eqref{nearHagnu} with parameters given in \eqref{AdS4coeff} and constant $\tilde g^{(\mathcal M)}=R^2_{\mathcal M}=4$.
At NLO we have the three-dimensional oscillator in $r$ (related to the $AdS_4$ factor) and a one-dimensional oscillator in $z$ with frequency
\be 
\omega_z^2 =  \frac{2}{\alpha'} \frac{4\,\kappa^2}{\left(1-4\,\kappa^2 \right)}\,.
\ee 
Pushing the computation at NNNLO we find
\be 
\beta_H^2 = \frac{8\pi^2 \alpha'}{1- 4\kappa^2} \left( 1 + c_0 \, \alpha'^{\frac12} + \tilde c_1 \, \alpha'\ + \tilde c_2 \, \alpha'^{3/2} + ...\right) 
\,,
\ee
with
\begin{align}
    &c_0= -\frac{3 + 2|\kappa|}{\sqrt{2}\sqrt{1-4\,\kappa^2}}\,,\\
    &\tilde c_1=\frac{5-2|\kappa| \left( 6 - 10 |\kappa| \right)}{4 \left(1-4\kappa^2 \right)} +\frac{2\pt\log2}{1-4\,\kappa^2}\bigl(3+8\,\kappa^2 \bigr)\,, \\[1.5ex]
    &\tilde c_2=-\frac{8-2|\kappa| \{ 83 - 2|\kappa|\left[14+2|\kappa| \left(120-2|\kappa| (78-70 |\kappa|) \right) \right] \}}{2^6 \sqrt{2} |\kappa|\left(1-4\kappa^2 \right)^{3/2}}+\nonumber\\
&\qquad\qquad\qquad\qquad\qquad\qquad\qquad - \frac{3\pt\log2}{\sqrt{2}(1-4\pt\kappa^2)^{3/2}}\bigl(3+2|\kappa|\bigr)\bigl(3+8\pt\kappa^2\bigr)\,.
\end{align}

\section{Conclusions}
\label{sec:conclu}
In this paper we have presented a formula (equation \eqref{implicitbetaH} equipped with \eqref{gendeltaE} and \eqref{newDelta}) allowing for the calculation of the Hagedorn temperature $T_H$ of rotating strings in backgrounds dual to confining theories (including CFTs on spheres).
Its main message is that the sigma-model analysis is sufficient to obtain the result up to NLO in the $\alpha'$ expansion, and it provides a contribution at NNLO which is not calculable (up to now) in the effective approach (the thermal scalar description).
Taking also the latter approach into account, we have provided a complete formula for $T_H$ at NNNLO (in the case of vanishing NSNS $B$ field), along the lines of \cite{Bigazzi:2023hxt, Bigazzi:2024biz, Canneti:2025cos}.   
The rotation of the string can describe, in the dual QFTs, the spins or some global charges, such as the R-charge, depending on the planes of rotation.

The sigma-model analysis entails the quantization of the quadratic fluctuations of the rotating string in the relevant geometry, written in the comoving frame. 
The equation for the fluctuating modes is technically more involved than in the non-rotating case, including now the pullback on the world-sheet of the Christoffel symbols of the target space.
Apart from deforming in a specific way the masses of the world-sheet modes, the rotation can thus induce some shifts of the frequencies.
The zero-point energy of the string, which includes $\log 2$ factors at NNLO, allows for the calculation of the Hagedorn temperature as the one at which the string ground state becomes massless.

At NNLO the analysis of the quadratic fluctuations is not sufficient to capture the complete result, which would require to calculate quartic terms in the fluctuations.
Although this is doable in principle, it is quite complicated in practice. 
Luckily enough, the effective approach can be easily employed to calculate the missing pieces at NNLO and NNNLO, while itself missing the $\log 2$ factors mentioned above. 

We have applied our formula in a number of examples and for different rotations, complementing the existing results in the literature \cite{Seitz:2025wpc,Ekhammar:2025efc}.\footnote{In the flat space case, both the sigma model and the effective approaches give the exact result.}
In particular, we have calculated the spin and R-charge dependence of the Hagedorn temperature in theories dual to global $AdS$ spaces, including ${\cal N}=4$ SYM, the ABJM theory, $AdS_3$ duals (also with rotations in the internal torus), and along the Reeb vector of general Sasaki-Einstein manifolds. 
For the truly confining case we have reported the explicit example of the WYM theory.
Of course, other confining theories can be treated with the same general formula in a straightforward way.

\vskip 15pt \centerline{\bf Acknowledgments} \vskip 10pt 

\noindent 
This project has been partially supported by the grant PRIN 20227S3M3B “Bubble Dynamics in Cosmological Phase Transitions”. The work of T.C. is supported by the Simons Foundation grant  994300 (``Simons Collaboration on Confinement and QCD Strings”). The work of F.C. is supported by the DFG Research Unit 5582 (“Modern foundations of Scattering Amplitudes"), project number 508889767.


\appendix

\section{Physical intuition about rotating strings}
\label{app:flatspace}

Let us consider a (type II) string which rotates on some two-dimensional conformally flat submanifold of the target space, with (imaginary) angular velocity. To fix ideas, let us assume that the plane of rotation is generated by $x^1$ and $x^2$. Then, we can switch to a comoving reference frame $\{q^1,q^2\}$ as\footnote{Let us stress again that we are dealing with Euclidean time. Therefore, $\omega$ is a real quantity associated with the imaginary angular velocity $\Omega=i\pt\omega$ of the string. See footnote \ref{imaginaryfoot} for a discussion about it.}
\be\label{firstcomoving}
x^1 + i \, x^2 = \(q^1 + i \,  q^2 \) \hspace{1pt} e^{- i \hspace{1pt} \omega  x^0} \, .
\ee
If, at the quantum level, $x^1$ and $x^2$ satisfy massive Klein-Gordon equations with mass $m$, namely
\be\label{eomX}
\( -\eta^{\alpha\beta} \partial_\alpha \partial_\beta + m^2 \) x^i = 0 \, , \quad i=1,2 \, ,
\ee
we may ask what the solutions for $q^1$ and $q^2$ are once $x^0$ is evaluated on the Hagedorn configuration \eqref{Hagreg}.

It follows that the on-shell comoving directions are constrained by
\be\label{eomcomovingframe}
-\eta^{\alpha\beta} \partial_\alpha \partial_\beta \, q +2\pt i \, A \cdot \partial_\sigma q + \( m^2 + \nu^2 \) q = 0 \, , 
\ee
where
\be
q = \begin{pmatrix}
    q^1 \\ q^2
\end{pmatrix} \, , \quad
A=\begin{pmatrix}
    0 & +i\pt\nu \\ -i\pt\nu & 0
\end{pmatrix} \, , \quad
2\pi\nu=\beta_H \omega \, .
\ee
With
\be
\xi = U^{-1} \cdot q \, , \quad U = \begin{pmatrix}
    i & -i \\ 1 & 1
\end{pmatrix} \, ,
\ee
the above equations can be reduced to the following diagonal form
\be\label{diagonalizedflateom}
-\eta^{\alpha\beta} \partial_\alpha \partial_\beta \, \xi + 2\pt i \, \mathcal A \cdot \partial_\sigma \xi + \( m^2 + \nu^2 \) \xi = 0 \, , \quad \mathcal A = \begin{pmatrix}
    \nu_1 & 0 \\ 0 & \nu_2 
\end{pmatrix}  \, ,
\ee
where
\be
\nu_1 = + \nu \, , \quad \nu_2 = - \nu \, .
\ee
Notice that this result shares the very same structure with the general equations of motion in \eqref{diagonalizedgeneom}.

For this discussion, the mode expansion in \eqref{genbosonicmodes} for $b=1,2$ can be understood as the ansatz for the solutions to \eqref{diagonalizedflateom}, for some frequencies $\omega_{bn}$ such that $\omega_{b,-n}=-\omega_{bn}$ and imposing
\be
\bigl[\alpha^b_n , \alpha^{b'}_m\bigr] = \bigl[\tilde\alpha^b_n , \tilde\alpha^{b'}_m\bigr] = \omega_{bn} \pt \delta^{bb'} \pt \delta_{m+n,0} \, , \quad \[a^b, \bigl(a^{b'}\bigr)^\dagger\] = \delta^{bb'} \, .
\ee
Notice that the presence of the zero-modes and the normalization adopted for the whole solution are crucial to satisfy the canonical commutation relations
\be
\bigl[\xi^b(\tau,\sigma), \partial_\tau \xi^{b'}(\tau,\sigma') \bigr] = 2\pi i \alpha' \delta^{bb'} \delta(\sigma-\sigma') \, .
\ee
Moreover, the expressions in \eqref{genbosonicmodes} are such that $\xi(\tau,\sigma+2\pi)=\xi(\tau,\sigma)$. Of course, the same applies to the comoving directions $q^1$ and $q^2$. This automatically realizes the boundary conditions introduced in (2.21) of \cite{Seitz:2025wpc}.

What is missing here is just an explicit formula for the frequencies. One can check that the equations in \eqref{diagonalizedflateom} are solved by the same expressions for $\omega_{b0}$ and $\omega_{bn}$ as the ones given in \eqref{genericshiftedharmonics}, for $b=1,2$ and $\mu_1=\mu_2=m$. We conclude that the net effect of the rotation on the comoving coordinates is to shift the standard harmonics as expected, without changing the values of the corresponding masses. Notice that, in the $m\to0$ limit, we recover the flat space results discussed in \cite{Seitz:2025wpc} (see also references therein).
Of course, the discussion can be generalized to several planes of rotation. 

\subsubsection*{Example: spinning strings in flat space}

To fix ideas, we are going to review the ten-dimensional flat space case from our perspective. So, let us consider a rotating string in $n_f$ orthogonal two-planes within
\be
\rmd s^2 = \(\rmd x^0\)^2 + \rmd \vec x^{\hspace{1pt}2} \, ,
\ee
for $n_f\leq4$. In the following, $\omega_k$, $k=1,\ldots,n_f$, will be associated with the (imaginary) angular velocities related to each plane of rotation. 

It is convenient to define the comoving coordinates as
\be\label{flatcomoving}
x^{2k-1} + i \, x^{2k} = q_k  \, e^{i(\phi_k-\omega_k x^0)} \, , 
\ee
so that 
\be
\rmd s^2 = \(\rmd x^0\)^2 +  \rmd \vec y^{\, 2} + \sum_{k=1}^{n_f} \Bigl(\rmd q_k^2 + q_k^2 \(\rmd \phi_k - \omega_k \rmd x^0\)^2\Bigr)  ,
\ee
where
\be
\label{y_coord}
y^j = x^{j+2\pt n_f} \, , \quad j=1,\ldots,9-2 \pt n_f \, .
\ee
Notice that, in these coordinates, the classical string configuration is static and in the Hagedorn regime reduces to \eqref{Hagreg}. Moreover, the resemblance of the above metric with the twisted backgrounds in \eqref{boostedgenericmetric} is manifest.

Needless to say, we are still in flat space. Consistently, the bosonic mass matrix defined in \eqref{bosmassmatr} is null in this scenario. As a consequence, the zero-point energy defined in \eqref{newDelta} reduces identically to\footnote{Remember that each frequency appears twice in the sum over all the bosonic directions (cf.~\eqref{firstcomoving}).}
\be
    \label{Delta_flat}
\Delta = 1 - \sum_{k=1}^{n_f} |\nu_k| \, , \quad 2\pi\nu_k = \beta_H \pt \omega_k \, .
\ee
On the other hand, $\Delta \mathcal E$ is exactly zero in this case (see section \ref{sec:effflatspace}). All in all, the Hagedorn temperature is simply fixed by
\be\label{flatrotatingbetaH}
\frac{\beta^2_H}{8\pi^2\alpha'} = 1 - \sum_{k=1}^{n_f} |\nu_k| \, ,
\ee
which is the direct generalization of formula (2.32) of \cite{Seitz:2025wpc} to any number of (imaginary) angular velocities. Let us recall that we have obtained it working with the Green-Schwarz (GS) formulation of Type II superstring theory. Finally, let us stress that the result \eqref{flatrotatingbetaH} is exact.


\section{Computation of the world-sheet zero point energy}
\label{compdelta}

Let us define
\be
E_{\nu}(\mu;s) = \sum_{n=1}^\infty \[(n+\nu)^2 + \mu^2\]^{-s/2} \, ,
\ee
for some generic parameters $\mu$ and $\nu$, and
{\small{
\be\label{regDelta}
\Delta(s) + \frac12 \sum_{b=1}^8 \omega_{0b} = -\frac12 \sum_{b=1}^8 \Bigl[ E_{\nu_b}(\mu_b;s) + E_{-\nu_b}(\mu_b;s) \Bigr] + \frac12 \sum_{f=1}^8 \Bigl[ E_{-\frac12+\nu_f}(\mu_f;s) + E_{-\frac12-\nu_f}(\mu_f;s)\Bigr] ,
\ee
}}where $\omega_{0b} = \sqrt{\mu^2_b + \nu^2_b}$. Clearly, it holds that
\be\label{removingregulator}
\Delta = \lim_{s\to-1} \Delta(s) \, ,
\ee
where $\Delta$ is the zero-point energy defined in \eqref{genDelta}. Thus, the strategy consists in rewriting each of the above terms so as to make their divergent behavior in the $s\to-1$ limit manifest. A proper cancellation will ensure that \eqref{removingregulator} is well-defined.

Notice that the world-sheet masses are expected to be small in the Hagedorn regime. This is clear from the definitions in \eqref{bosonictrace}, which make them linear in $\beta_H\sim {\cal O}\mathcal(\sqrt{\alpha'})$ by construction.
Therefore, we can expand equation \eqref{regDelta} in powers of the bosonic world-sheet masses. Useful formulae are the definition of the (generalized) Riemann Zeta functions
\be
\zeta(s,a) = \sum_{n=0}^\infty (n+a)^{-s} \, , \quad \zeta(s) = \sum_{n=1}^\infty n^{-s} = \zeta(s,1) \, ,
\ee
and well-known relations among Gamma functions that give
\be
\Gamma(s/2+k)\Gamma(1-s/2-k) = (-1)^k \, \Gamma(s/2) \Gamma(1-s/2) \, .
\ee
They provide
\be
\frac{\partial^k E_\nu (\mu, s)}{\partial (\mu^2)^k}  \biggr |_{\mu=0} = \frac{\Gamma(s/2+k)}{\Gamma(s/2)} (-1)^k \zeta(s+2k, \nu+1) \, ,
\ee
and so
\begin{subequations}
\begin{align}
& E_{\nu} (\mu,s) = \zeta(s,1+\nu) - \frac12 s \, \zeta(s+2,1+\nu) \, \mu^2 + \mathcal O(\mu^4) \, ,\\
&E_{-1/2+\nu} (\mu,s) = \zeta(s,1/2+\nu) - \frac12 s \, \zeta(s+2,1/2+\nu) \, \mu^2 + \mathcal O(\mu^4) \, .
\end{align}
\end{subequations}
The latter are exactly the functions appearing in \eqref{regDelta}.

We can now massage each term in the expansion using well-known
properties of the Zeta functions, namely
\be\label{zetaformulae}
\zeta(-1,a) = -\frac{1}{12} + \frac12 a - \frac12 a^2 \, , \quad \zeta(s+2,a) = \frac{1}{s+1} - \psi(a) + \mathcal O(s+1) \, ,
\ee
for some complex number $a$ with $\text{Re}\pt a  >0$,\footnote{If $\nu$ is imaginary, then the condition for the validity of the formulae in \eqref{zetaformulae} is automatically satisfied. Furthermore, if $\nu$ is real, then everything is fine as long as $|\nu|<1/2$, see also \cite{Seitz:2025wpc}.} $\psi$ being the digamma function. The former follows from, \eg \cite[\href{https://dlmf.nist.gov/25.11.E14}{(25.11.14)}]{NIST:DLMF} and \cite[\href{https://dlmf.nist.gov/24.2.T2}{Table 24.2.2}]{NIST:DLMF}), while the latter has been derived in \cite{ec8cdabf-0f03-3b34-9f69-879792833dbb}. Moreover, let us stress that the cancellation of the conformal anomaly requires that
\be
\sum_{b=1}^8 \mu^2_b = \sum_{f=1}^8 \mu^2_f \, . 
\ee
Its origin and validity in the Hagedorn regime is deeply discussed in \cite{Bigazzi:2024biz, Canneti:2025cos}, where we referred to it as \emph{mass-matching condition}. Moreover, it holds true even in any scenario relevant to this paper. This is precisely what guarantees the cancellation of the pole in the Laurent expansion above.

All in all, it follows that the $s\to-1$ limit of \eqref{regDelta} reads
\be
\begin{split}
    \Delta = 1 - \frac12  \sum_b & \omega_{0b} + \frac12\biggl(\sum_b \nu_b^2 - \sum_f \nu_f^2 \biggr) +\\ &+\frac14\sum_b \Psi_1(\nu_b) \, \mu^2_b - \frac14\sum_f \Psi_{1/2}(\nu_f) \, \mu^2_f + \mathcal O (\mu^4_b) + \mathcal O (\mu^4_f)\, ,
\end{split}
\ee
where we defined
\be
\pt\Psi_a(\nu) = \psi(a+\nu) + \psi(a-\nu) \, .
\ee
In any scenario discussed in this work, the angular velocities $\omega_{b,f}$ related to massive modes have been considered fixed in the $\alpha'$ expansion. This makes $2\pi\nu_{b,f}=\beta_H \pt \omega_{b,f}$ as small as $\beta_H$ or $\mu_{b,f}$. Since
\begin{subequations}
    \begin{align}
        &\Psi_1(\nu_b) = -2\pt \gamma_E + \mathcal O (\nu_b^2) \, , \\
        &\Psi_{1/2}(\nu_f) = - 2\pt\gamma_E -4\log 2 + \mathcal O (\nu_f^2) \, ,
    \end{align}
\end{subequations}
$\gamma_E \approx 0.577$ being the Euler constant, we conclude that the above expression for $\Delta$ gives the zero-point energy in \eqref{newDelta} on the shift-matching condition \eqref{shiftmatch}.

\section{Alternative formulation}
\label{app_geom}
Let us now rephrase what we discussed in appendix \ref{app:flatspace} in terms of geometrical objects that we introduced in section \ref{sec:gen_proposal}. To do so, we will also consider the simple examples of a spinning string with a single angular velocity in both flat space and global $AdS$.

The ten bosonic fluctuations ($\zeta$) around the world-sheet minimal configuration can be decomposed into eight normal ($\chi$) and two longitudinal ($\xi$) modes
\begin{equation}
    \zeta^p = x^p_\alpha \xi^{\alpha} + N^p_{i} \chi^{i}\,,
\end{equation}
where we denote with $ x^p_\alpha$ and $N^p_{i} $ ($i=1,\cdots,8$) the tangent and normal vectors to the world-sheet, respectively.\footnote{Notice that within a world-sheet-adapted frame basis for the ambient spacetime, we can identify $N^p_{i}$ with the vielbein along the transverse direction with respect to the world-sheet.} Remarkably, only the $\chi$ modes have a physical relevance, since the longitudinal $\xi^\alpha$ are associated with world-sheet diffeomorphism and, in principle, can be, in turn, gauge fixed to zero. More details on the above quantities can be found in e.g. \cite{Forini:2015mca,Bigazzi:2023hxt,Bigazzi:2023oqm,Bigazzi:2024biz,Singh:2023olv}.

In analogy with (\ref{bosmassmatr}), the masses of the physical modes $\mu_b$ can be computed as the eigenvalues of the bosonic mass matrix $\mathcal M_b$, whose components are\footnote{For the sake of simplicity, we set the Kalb-Ramond field of the supergravity background to zero. Moreover, in all the cases of interest for this work, the so-called extrinsic curvature $K^i_{\alpha\beta}$, and not just its trace, vanishes identically. We recall that the (Nambu-Goto) string minimal classical configuration $X^m$ is such that $h^{\alpha\beta}K^m_{\alpha\beta}=0$~\cite{Forini:2015mca}.}
\be\label{mqui}
\( \mathcal M_b \)_{ij} = -\eta^{\alpha\beta} \partial_\alpha X^p \partial_\beta X^q R_{m p n q} (X)N^m_iN^n_j\,,
\ee
where $\eta_{\alpha\beta}$ is the two-dimensional Minkowski metric. Let us stress that again $\sigma^\alpha = \{\tau, \sigma\}$ denotes the coordinates on the world-sheet, while $R_{mp n q}$ is the Riemann tensor of the target space. Furthermore, in writing down (\ref{bosmassmatr}), (\ref{mqui}), we have chosen the so-called conformal gauge for the auxiliary metric, namely $h_{\alpha\beta} = e^\Lambda \eta_{\alpha\beta}$.  Moreover, notice that now the sum of the masses is given by 
\be
\Tr \( \mathcal M_b \) = -\eta^{\alpha\beta} \partial_\alpha X^p \partial_\beta X^q R_{m p n q}(X)N^m_iN^n_j\delta^{ij} = \sum_{b=1}^8 \mu^2_b \, .
\ee
Embracing the same assumptions of section \ref{sec:gen_proposal}, namely linear embedding maps, zero extrinsic curvature, and a vanishing Kalb-Ramond field, the on-shell bosonic fluctuations turn out to be solutions of
\be
\label{EqQ}
-\eta^{\alpha\beta} D_\alpha D_\beta \, \chi + \mathcal M_b \cdot \chi  = 0 \, ,
\ee
where now the covariant derivative $D$ is defined using the normal bundle connection, i.e.
\begin{equation}
\label{EqA}
    D_\alpha \chi^i = \partial_\alpha \chi^i -i A_{\alpha}{}^i_j\,\chi^j\,,\quad A_{\alpha\,ij} =  -i\,\partial_\alpha X^p \, \Gamma^m{}_{pq} \, g_{mn} N^n_iN^q_j  \,.
\end{equation}
Remarkably, the above-defined connection in (\ref{EqA}) is what keeps trace of the background field related to the chemical potential that we switched on. The main difference of this $SO(8)$ connection with that in (\ref{AChristoffels}), is that the latter could display some singular behavior in looking at spinning string configurations placed along contractible cycles. This issue is cured by the splitting of the fluctuations and the introduction of $A_{\alpha\,ij}$.

Let us now provide some useful examples.

\subsection{Flat spacetime}
We start by focusing on a string in a twisted flat spacetime, whose metric is given by \cite{Seitz:2025wpc}
\begin{equation}
\label{Eq: flat metric}
		\rmd s^2 =  \,\left(\rmd x^0\right)^2+\rmd q^2  +  q^2\left(\rmd \phi -\omega \rmd x^0\right)^2  + \rmd \vec y^{\, 2}\,,
    \end{equation}
with the $y^i$ coordinates defined as in (\ref{y_coord}). Here we can take a string embedding as\footnote{Clearly, we should set the string at the position $q=0$ once we have evaluated all the geometrical quantities, not to lose information. In another way, we should set $q = \epsilon$, to evaluate all the geometrical objects and then send $\epsilon\to 0$. }
\begin{equation}
     t = \rho \sigma\,,\quad y_1 = J \tau\,, \quad \phi = \text{const.}\, \quad q = 0\,, \quad y_{i} = \text{const.}\,, \,\, i = 2,\cdots, d-3\,,
\end{equation}
where in the Hagedorn regime 
\begin{equation}
    \rho_H = \frac{\beta_H}{2\pi}\,,\quad J =0\,.
\end{equation}
A vanishing extrinsic curvature characterizes this precise string configuration. The tangent vectors to the world-sheet here are given by
\begin{equation}
    x^m_\tau = \left(0,0,0,1,0,0,0,0,0,0\right)\,,\qquad  x^m_\sigma = \left(1,0,0,0,0,0,0,0,0,0\right)\,,
\end{equation}
and the normal ones are
\bea
\label{Nflat}
&&N_1^m = \left(0,1,0,0,0,0,0,0,0,0\right)\,,\quad N_2^m = \left(0,0,\frac{1}{q},0,0,0,0,0,0,0\right)\,,\nonumber\\
&& N_i^m = \left(0,0,0,0,0,\cdots, 1,\cdots,0\right)\,, \quad i = 2, \cdots, d-3\,.
\eea
Thus, by looking at the connection in the normal bundle, we have that the following components are non-zero (in the Hagedorn regime):
\bea
\label{Aij_flat}
A_{\sigma\,ji} &= -i \rho \Gamma^{m}_{0 p}N^p_iN^s_jg_{ms} = -i \frac{\rho}{2}N^p_iN^s_j\left(\partial_pg_{s 0}-\partial_s g_{p0}\right)\,,\notag\\
&=  -i\left(\delta_{j1}\delta_{i2}-\delta_{j2}\delta_{i1}\right)\omega \rho_H\,.
\eea
In the $(q,\phi)$ subspace the $A_{\sigma}$ connection can be diagonalized as
\bea
\label{ConnectionAsigma_flat}
\mathcal{A}= U^{-1} A_\sigma U = \omega\rho_H\, \sigma_3\,,
\eea
where also the fluctuations are rotated as $\chi = Uy$\,. We can  thus write down the equations of motion (notice again that in flat spacetime, all the mode masses are zero), namely
\be
(\partial^2_\tau -\partial^2_\sigma) \, \chi + 2 \hspace{1pt} i \,  \mathcal{A} \hspace{-2pt} \, \partial_\sigma \, \chi + \mathcal{A}^2 \, \chi = 0 \, ,
\ee
that can be solved as in section \ref{sec:gen_proposal}, and thus providing an analogous result for the zero-point energy (and in turn for the Hagedorn temperature) as in \eqref{Delta_flat}.

\subsection{Global $AdS$}
The twisted Euclidean global $AdS_5\times S^5$ background can be expressed as
	\begin{equation}\label{Eq: 10d metric}
		\rmd s_{10}^2 = \cosh^2 r \,\rmd t ^2+\rmd r^2  + \sinh^2 r \,\rmd \tilde\Omega_3^2 + \rmd \Omega^2_5\,,
	\end{equation}
	where the $\rmd \Omega^2_5$ metric of the internal $S^5$ and the twisted $S^3$ metric are given as follows 
	\begin{equation}
	\begin{aligned}
        & \rmd \Omega_5^2 = \rmd \theta^2 +\cos^2\theta\rmd \phi^2 + \sin^2\theta \rmd \Omega^2_3\,,\notag\\
		&\rmd \tilde\Omega_3^2 = \rmd \alpha_1^2 + \sin^2\alpha_1(\rmd \alpha_2^2 + \sin^2\alpha_2 (\rmd \alpha_3 -\omega \rmd t)^2)\,.
	\end{aligned}
	\end{equation}
The type IIB supergravity solution is completed by the self-dual five-form flux
	\begin{equation}\label{five_flux}
		F_5 = \left(1+ \star\right)G_5\,,\quad G_5 =  - 4 \sinh^3 r \cosh r\,\rmd \tau \wg \rmd r\wg \text{vol}_{\tilde S^3}\,.
	\end{equation}
Here, we have set the $AdS$ radius $\ell$ to one.\footnote{Moreover, here we are using $\alpha^\prime,g_s= 1$ units.}

In this case, the embedding of the rotating string can be taken as 
\begin{equation}
    t = \rho \sigma\,,\quad \phi = J \tau\,,\quad \rho = \frac{\beta}{2\pi}\,.
\end{equation}
More specifically, the string wraps the Euclidean time  $\tau$ and the angular $\phi$ isometry inside the five-sphere.\footnote{One in principle could freely choose also other $U(1)$ isometries inside $S^5$, since in the Hagedorn limit, we will set all the angular momenta to zero.} 
In particular, the induced metric is given by
\begin{equation}
\label{ind_string}
\rmd s_{\text{2}}^2 = \rho^2(\cosh^2r +\omega^2 \sinh^2 r\,\sin^2\alpha_1\, \sin^2\alpha_2  )\rmd \sigma^2 + J^2\cos^2\theta\rmd \tau^2 \,. 
\end{equation}
To minimize the action, we impose the string to localize at the “center” of $AdS$ $r =0$, and fix $\theta = 0$. Hence, the induced metric in equation (\ref{ind_string}) reduces to 
\begin{equation}
\label{ind_string_2}
\rmd s_{2}^2 = J^2 \rmd \tau^2 + \rho^2\rmd \sigma^2\,. 
\end{equation}
Then, it is possible to verify that the present configuration is further characterized by vanishing extrinsic curvature.

By expanding the string action at the quadratic order in the scalar fluctuations, we can derive the following mass matrix for the transverse modes ($\chi$)
\be
\( \mathcal M_b \)_{ij} = -\eta^{\alpha\beta}R_{\alpha i \beta j}\,,
\ee
that in the Hagedorn limit is given by
\begin{equation}
\label{M_globalAdS}
\mathcal M_b = \mu^2 \, \text{diag}\{\mathds 1_4, \mathbb 0_4\}
\, , \quad \mu = \frac{\beta_H}{2\pi} \, .
\end{equation}
In this case, the metric is twisted in the $AdS$ sector of the background, so only its fluctuations will see the arising of a normal bundle connection in the covariant derivative.

The “$AdS$” tangent vector to the world-sheet is given by
\bea
x_\sigma^m = \left(1,0,0,0,0,0,0,0,0,0\right)\,,
\eea
while it is natural to define the normal vectors inside $AdS$ as
\bea
\label{Nads}
&&N_1^m = \left(0,\frac{1}{\sinh r},0,0,0,0,0,0,0,0\right)\,,\quad N_2^m = \left(0,0,\frac{1}{\sin\alpha_1\sinh r},0,0,0,0,0,0,0\right)\,,\nonumber\\
&& N_3^m = \left(N_3^0,0,0,N_3^{\alpha_3},0,0,0,0,0,0\right)\,,\quad N_4^m = \left(0,0,0,0,1,0,0,0,0,0\right)\,.
\eea
The coefficients $N_3^{0\,,\alpha_3}$ can be fixed by requiring
\begin{equation}
\label{embed:ortho}
  N_i^m x_\alpha^n g_{mn} =0\,,\quad 
  N_i^m N_j^n g_{mn} = \eta_{ij}~, \quad i\,,j= 1,2,\ldots , 6\,,
\end{equation}
from which we obtain
\bea
\label{NtN3}
N_3^{0} = -\frac{g_{0\alpha_3}}{\sqrt{g_{00}}\sqrt{\det\tilde g}}\,,\quad N_3^{\alpha_3} = \frac{\sqrt{g_{00}}}{\sqrt{\det\tilde g}}\,,\quad \det\tilde g = g_{00}g_{\alpha_3\alpha_3}- g_{0\alpha_3}^2\,.
\eea
It is possible to verify that the normal vectors of (\ref{Nads}) and (\ref{NtN3}) verify the completeness relation 
\begin{equation}
\label{embed:complete}
	g^{\alpha\beta} x_\alpha^m x_\beta^n + \eta^{ij} N_i^m N_j^n = g^{mn}~.
\end{equation}
Hence, we can calculate the connection  $A_{\sigma \,ij}$ (relevant for the $AdS$ sector): its non-vanishing components (at the center of $AdS$) are given by
\begin{align}
\label{A43}
&A_{\sigma43} =-A_{\sigma34} = \frac{1}{2}\left(N^0_3\partial_rg_{00}+N^{\alpha_3}_3\partial_rg_{\alpha_3 0}\right) = i\omega\text{s}_1\text{s}_2\,,\notag\\
& A_{\sigma13} =-A_{\sigma31}= \frac{1}{2\sinh r}\left(N^0_3\partial_{\alpha_1}g_{00}+N^{\alpha_3}_3\partial_{\alpha_1}g_{\alpha_3 0}\right) = i\omega \text{c}_1\text{s}_2\,,\nb\\
&A_{\sigma 23} =-A_{\sigma32}= \frac{1}{2\text{s}_1\sinh r}\left(N^0_3\partial_{\alpha_1}g_{00}+N^{\alpha_3}_3\partial_{\alpha_1}g_{\alpha_3 0}\right) = i\omega\text{c}_2\,,
\end{align}
where $\text{c}_{1,2} = \cos\alpha_{1,2}$, and $\text{s}_{1,2} = \sin\alpha_{1,2}$.

Finally, it is possible to diagonalize the above connection via a fluctuation rotation, obtaining
\begin{equation}
    \mathcal{A} = A_\sigma = \omega\, \text{diag}\{\sigma_3, \mathbb 0_6\}\,.
\end{equation}
Let us notice that the mass matrix in \eqref{M_globalAdS} is not affected by the modes' rotation. Moreover, it is clear that 
\begin{equation}
    \left[\mathcal{A}, \mathcal{M}_b\right] =0\,.
\end{equation}
It is then possible to generalize this analysis to the case of multiple angular velocities, as in \cite{Ekhammar:2025efc} and in the main body of the present work. This case is just computationally more complicated with respect to the one above, but the results are similar.

\section{The coupling to $R$-charges from a string perspective}
\label{app:Rcharged}
In the main body we dealt with several twisted metrics on spheres in the comoving frame of the string. For instance, in section \ref{sec:spinonSn}, we either twisted some stereographic directions or one of the toroidal coordinates on the sphere (cf.~\eqref{boostedsphere} with \eqref{toroidalsphere},\eqref{chargedshift}). The two scenarios lead to two completely different world-sheet spectra, and so to different predictions about the corresponding Hagedorn temperature. 

It would be interesting to translate such differences into different proper motions of the string. Of course, the metric \eqref{boostedsphere} corresponds to a rotating string in several conformally flat submanifolds of the sphere. What about the shift in \eqref{chargedshift}? The relations that give the polar angle of a sphere in terms of its toroidal coordinates has been given in \eqref{toroidalcoordinate}. One can check that the inverse transformations are
\be
\psi=\arctan (\cos\theta, \,\sin\theta\cos\tilde\theta) \, , \quad \phi = \arccos(\sin\theta\sin\tilde\theta) \, .
\ee
Then, for small values of $\kappa$,\footnote{It is the same regime as in section 3.1.1 of \cite{Ekhammar:2025efc}.} we have
\be
\theta \mapsto \theta + i \pt  \cos\tilde\theta \,\kappa \, t + \mathcal O\(\kappa^2\)\, .
\ee
According to \eqref{stereocoord}, this translates into a transformation of the stereographic radius in the same regime. All in all, the change of coordinates in \eqref{chargedshift} reduces to a shift of the radial stereographic coordinate that is not affected in the twisted metric \eqref{boostedsphere}. For this reason, some of the world-sheet bosons get a $\kappa$-dependent mass (see \eqref{chargedspectrum}), in contrast to the purely rotating case.

Now, let us focus on the five-dimensional sphere. In section \eqref{sec:reeb}, we interpreted $\kappa$ as the chemical potential that couples to one of the R-charges of planar $\mathcal N=4$ super Yang Mills at finite temperature. The same understanding applies to the parameter $\kappa$ that appears in the map \eqref{chargedshift}. However, it must be related to a different $U(1)$ subgroup of the entire $SO(6)$ R-symmetry group. Indeed, it gives rise to a different world-sheet spectrum (cf.~\eqref{chargedspectrum} to \eqref{SEspectrum}).
Thus, we may wonder how the shift in \eqref{chargedshift} affects the round metric of the sphere when expressed in the coordinates of section \ref{sec:reeb}.

The static metric on $S^5$ as a $U(1)$ bundle over $\CP^2$ reads (see e.g.~\cite{Gauntlett:2004yd})
\begin{align}\label{NOTboostSEmetricS5}
    \rmd s^2_{S^5} =  &\left[\rmd\psi' -\frac12 \sin^2\chi \(\rmd \alpha_3 + \cos\alpha_1 \rmd \alpha_2 \) \right]^2 + \rmd \chi^2 + \\
    &+\frac14 \sin^2\chi \(\rmd\alpha_1^2+\sin^2\alpha_1\rmd\alpha_2^2\)+ \frac14 \sin^2\chi \cos^2\chi \(\rmd\alpha_3+\cos\alpha_1\rmd\alpha_2\)^2 \, , \nonumber
\end{align}
with $\psi'$ ($\alpha_3$) having period $2\pi$ ($4\pi$). In general, the metric on $S^5$ can be obtained embedding the sphere in $\mathds C^3$, that is
\be\label{S5inC3}
\rmd s^2_{S^5} = |\rmd z_1|^2 + |\rmd z_2|^2 + |\rmd z_3|^2 \, , \quad | z_1|^2 + | z_2|^2 + | z_3|^2 = 1 \, .
\ee
The reader can check that the embedding map
\begin{subequations}\label{embedS5SE}
  \begin{empheq}[left=\empheqlbrace]{align}
    &z_1 = \sin\chi \, \cos(\alpha_1/2) \, \text{Exp}\[i\pt\psi'\] \, \text{Exp}\[-i\pt(\alpha_2+\alpha_3)/2\] \, ,\\[0.9ex]
    &z_2 = \sin\chi \, \sin(\alpha_1/2) \, \text{Exp}\[i\pt\psi'\] \, \text{Exp}\[+i\pt(\alpha_2-\alpha_3)/2\]   \, ,\\[0.75ex]
    &z_3=\cos\chi \, \text{Exp}\[i\pt\psi'\]\, ,
  \end{empheq}
\end{subequations}
leads to \eqref{NOTboostSEmetricS5}.

On the other hand, the metric on $S^5$ in toroidal coordinates \eqref{toroidalsphere} can be also computed in a similar way. In this case, the embedding map is
\begin{subequations}\label{toroidalembed}
  \begin{empheq}[left=\empheqlbrace]{align}
    &z_1 = \cos\phi \, \cos(t_1/2) \, \text{Exp}\[-i\pt(t_2+t_3)/2\] \, ,\\[0.9ex]
    &z_2 = \cos\phi\, \sin(t_1/2)  \, \text{Exp}\[+i\pt(t_2-t_3)/2\]   \, ,\\[0.75ex]
    &z_3=\sin\phi \, \text{Exp}\[i\pt\psi\]\, ,
  \end{empheq}
\end{subequations}
Here, $t_1$, $t_2$, $t_3$ are the Euler angles related to the Hopf fibration that describes the three-sphere in \eqref{toroidalsphere} as a $S^1$ bundle over $S^2$, namely (\eg see \cite{Betzios:2019rds})
\be
\rmd\Omega^2_3 = \frac14 \[\(\rmd t_3^2 + \cos t_1 \rmd t_2\)^2 + \rmd t_1^2 + \sin^2 t_1 \rmd t_2^2\] \, ,
\ee
where $0\leq t_1 < \pi$, $0 \leq t_2 < 2\pi$, $-2\pi \leq t_3 < 2\pi$.

Notice that the two sets of coordinates can be connected by equating each entry of the embedding map. Up to periodicities, we get
\begin{align}
    &\psi' = \psi + \pi \, , \quad \chi = \phi + \pi/2 \, , \\
    &\alpha_1=t_1\, , \quad \alpha_2 = t_2 \, , \quad \alpha_3=t_3 + 2\pt\psi + 2\pi \, .
\end{align}
As a consequence, the shift of the toroidal coordinate in \eqref{chargedshift} is not enough to realize the twist along the Reeb vector that leads to \eqref{SEmetricS5}. In fact, if we want only $\psi'$ to be affected, a compensating reparameterization of the fiber $S^1$ in $S^3$ is also necessary.

We can also extend the comparison to the metric of $S^5$ written in stereographic coordinates as in \eqref{stereometric}. The latter can be derived from \eqref{S5inC3} through the embedding map
\begin{subequations}
  \begin{empheq}[left=\empheqlbrace]{align}
    &z_1 = \frac{2\pt(\varphi_1 + i \pt \varphi_2)}{1+\varphi^2} = \frac{2\pt\eta_1 \, \text{Exp}\[i\pt\chi_1\]}{1+\varphi_5^2+\eta^2_1+\eta^2_2} \, ,\\[0.9ex]
    &z_2 = \frac{2\pt(\varphi_3 + i \pt \varphi_4)}{1+\varphi^2} = \frac{2\pt\eta_2 \, \text{Exp}\[i\pt\chi_2\]}{1+\varphi_5^2+\eta^2_1+\eta^2_2} \, ,\\[0.75ex]
    &z_3=\frac{2\pt\varphi_5 + i \pt (1-\varphi^2)}{1+\varphi^2}=\frac{2\pt\varphi_5 + i \pt (1-\varphi_5^2-\eta^2_1-\eta^2_2)}{1+\varphi_5^2+\eta^2_1+\eta^2_2}\, ,
  \end{empheq}
\end{subequations}
where $\varphi_1=\eta_1 \cos \chi_1$, $\varphi_2=\eta_1 \sin \chi_1$, $\varphi_3=\eta_2 \cos \chi_2$, $\varphi_4=\eta_2 \sin \chi_2$. Under these circumstances, we can have up to $n_S=2$ twists as in \eqref{boostedsphere} (for $p=0$, $d=4$). The latter act on $z_1$ and $z_2$, while keeping $z_3$ fixed. Therefore, it is conceptually orthogonal to what happens in \eqref{toroidalembed} under the shift in \eqref{chargedshift}.

All the scenarios discussed above can be resumed by the embedding map
\begin{subequations}
  \begin{empheq}[left=\empheqlbrace]{align}
    &z_1 = \mu_1 \, \text{Exp}\[i\pt\phi_1\] \, ,\\[0.9ex]
    &z_2 = \mu_2 \, \text{Exp}\[i\pt\phi_2\] \, ,\\[0.75ex]
    &z_3=\mu_3 \, \text{Exp}\[i\pt\phi_3\]\, ,
  \end{empheq}
\end{subequations}
where $\sum_i^3 \mu_i^2=1$ and $\phi_i\in[0,2\pi)$, $i=1$, $2$, $3$. With $\mu_i$, $i=1$, $2$, $3$, given in \eqref{mu1mu2mu3}, it produces the metric on $S^5$ presented in \eqref{metricS5inIR6}. It is clear how the shifts in \eqref{U13shifts} affect independently each complex variable $z_i$, $i=1$, $2$, $3$.

\bibliographystyle{utphys}

\providecommand{\href}[2]{#2}\begingroup\raggedright\endgroup

\end{document}